\documentclass[11pt,a4paper]{article}
\usepackage{jcappub}

\usepackage{epsfig}
\usepackage[english]{babel}
\usepackage{amsfonts,amsmath,amssymb}
\usepackage{enumerate}

\def\mg{m_{3/2}}   
\def\ms{m_\rm{s}}  
\def\Mm{M_\rm{m}}  
\def\Mnr{M_*}      
\def\Mp{M_\rm{P}}  
\def\fosc{\f_\rm{osc}}  %
\def\Hosc{H_\rm{osc}}  %
\def\Tosc{T_\rm{osc}}  %
\def\aD{ \alpha_{_D} } 
\def\mD{ \m_{_D} }     
\def\ZD{ Z_{_D}' }     
\def\aBL{\alpha_{_{B-L}}  }
\def\gBL{ g_{_{B-L}}  }
\def\MBL{ M_{_{B-L}}  }
\def\ZBL{ Z_{_{B-L}}' }

\def\dm{ \rm{DM}  } 

\def\v{ \rm{v}  } 
\def\d{ \rm{d}  } 
\def\dm{\text{\tiny DM}} 

\def\qX{ q_{_X} }
\def\nX{ n_{_X} }
\def\hX{ \eta_{_X} }

\def\Bv{B_\rm{v}}
\def\Lv{L_\rm{v}}
\def\BLv{(B-L)_\rm{v}}
\def\Bd{B_\rm{d}}

\def\gv{g_\rm{v}}
\def\gd{g_\rm{d}}

\def\gdN{g_\rm{d,BBN}}
\def\Tdec{T_\rm{dec}}
\def\Tv{T_\rm{v}}
\def\Td{T_\rm{d}}

\title{Affleck-Dine dynamics and the dark sector of pangenesis}

\author{Benedict von Harling, Kalliopi Petraki and Raymond R. Volkas} 
\affiliation{ARC Centre of Excellence for Particle Physics at the Terascale, School of Physics, \\ The University of Melbourne, Victoria 3010, Australia}

\emailAdd{bvo@unimelb.edu.au}
\emailAdd{kpetraki@unimelb.edu.au}
\emailAdd{raymondv@unimelb.edu.au}

\date{\today}

\abstract{Pangenesis is the mechanism for jointly producing the visible and dark matter asymmetries via Affleck-Dine dynamics in a baryon-symmetric universe. The baryon-symmetric feature means that the dark asymmetry cancels the visible baryon asymmetry and thus enforces a tight relationship between the visible and dark matter number densities.  The purpose of this paper is to analyse the general dynamics of this scenario in more detail and to construct specific models.  After reviewing the simple symmetry structure that underpins all baryon-symmetric models, we turn to a detailed analysis of the required Affleck-Dine dynamics. Both gravity-mediated and gauge-mediated supersymmetry breaking are considered, with the messenger scale left arbitrary in the latter, and the viable regions of parameter space are determined.  In the gauge-mediated case where gravitinos are light and stable, the regime where they constitute a small fraction of the dark matter density is identified.  
We discuss the formation of $Q$-balls, and delineate various regimes in the parameter space of the Affleck-Dine potential with respect to their stability or lifetime and their decay modes. We outline the regions in which $Q$-ball formation and decay is consistent with successful pangenesis.
Examples of viable dark sectors are presented, and constraints are derived from big bang nucleosynthesis, large scale structure formation and the Bullet cluster.  Collider signatures and implications for direct dark matter detection experiments are briefly discussed.  The following would constitute evidence for pangenesis: supersymmetry, GeV-scale dark matter mass(es) and a $Z'$ boson with a significant invisible width into the dark sector.}

\begin{document}
\maketitle
\newcommand{\nc}{\newcommand}
\def\cal{\mathcal}
\def\rm{\mathrm}

\nc{\eq}[1]{Eq.~(\ref{#1})}
\nc{\eqs}[1]{Eqs.~(\ref{#1})}

\nc{\pd}{\partial}
\nc{\bea}{\begin{eqnarray}}
\nc{\eea}{\end{eqnarray}}
\nc{\bal}{\begin{alignedat}}
\nc{\eal}{\end{alignedat}}
\nc{\beq}{\begin{equation}}
\nc{\eeq}{\end{equation}}
\nc{\bit}{\begin{itemize}}
\nc{\eit}{\end{itemize}}
\nc{\benu}{\begin{enumerate}}
\nc{\eenu}{\end{enumerate}}
\nc{\bdes}{\begin{description}}
\nc{\edes}{\end{description}}

\nc{\nn}{\nonumber}

\nc{\hc}{\rm{h.c.}}
\nc{\cc}{\rm{c.c.}}

\nc{\sub}[1]{_{\rm{#1}}}
\nc{\ssub}[1]{_{_\rm{#1}}}
\nc{\super}[1]{^{\rm{#1}}}
\nc{\ssuper}[1]{^{^\rm{#1}}}


\nc{\pare}[1]{\left( #1 \right)}
\nc{\sqpare}[1]{\left[ #1 \right]}
\nc{\ang}[1]{\langle #1 \rangle}
\nc{\abs}[1]{\left| #1 \right|}

\def\g5{\gamma_{5}}

\def \eV{\text{ eV}}
\def\keV{\text{ keV}}
\def\MeV{\text{ MeV}}
\def\GeV{\text{ GeV}}
\def\TeV{\text{ TeV}}

\def\erg{\: \rm{erg}}

\def \cm{\: \rm{cm}}
\def \km{\: \rm{km}}
\def \pc{\: \rm{pc}}
\def\kpc{\: \rm{kpc}}
\def\Mpc{\: \rm{Mpc}}
\def\Gpc{\: \rm{Gpc}}
\def \AU{\: \rm{A.U.}}

\def\sr{\: \rm{sr}}

\def \snd{\: \rm{s}}
\def  \yr{\: \rm{yr}}
\def \Myr{\: \rm{Myr}}
\def \Gyr{\: \rm{Gyr}}

\def \gr {\: \rm{g}}
\def \kgr{\: \rm{kg}}

\def\a{\alpha}
\def\b{\beta}
\def\g{\gamma}
\def\e{\epsilon}
\def\z{\zeta}
\def\h{\eta}
\def\th{\theta}
\def\i{\iota}
\def\k{\kappa}
\def\l{\lambda}
\def\m{\mu}
\def\n{\nu}
\def\ks{\xi}
\def\om{o}
\def\p{\pi}
\def\r{\rho}
\def\s{\sigma}
\def\t{\tau}
\def\y{\upsilon}
\def\f{\phi}
\def\x{\chi}
\def\ps{\psi}
\def\w{\omega}

\def\ve{\varepsilon}
\def\vr{\varrho}
\def\vs{\varsigma}
\def\vf{\varphi}

\def\G{\Gamma}
\def\D{\Delta}
\def\Th{\Theta}
\def\L{\Lambda}
\def\Ks{\Xi}
\def\P{\Pi}
\def\S{\Sigma}
\def\Y{\Upsilon}
\def\F{\Phi}
\def\Ps{\Psi}
\def\W{\Omega}


\section{Introduction}
\label{sec:intro}

The present-epoch matter content of our universe points to the existence of new particles and interactions beyond those described in the Standard Model (SM). Precision cosmological measurements have established that 4.6\% of our universe consists of visible matter (VM), and 23\% consists of dark matter (DM).

The known properties of the visible matter assert that its relic abundance today can be understood only if we hypothesise yet-unknown processes that created an asymmetry between the visible baryons and antibaryons in the early universe. The dynamical generation of such an asymmetry requires processes which violate baryon number, $C$ and $CP$ conservation, and occur out of equilibrium at earlier times. The survival of the baryonic asymmetry today implies that any baryon-violating processes must have been ineffective thereafter.

On the other hand, the properties of dark matter are poorly understood, and there are many possible mechanisms for the production of the DM relic density.
Well-studied DM scenarios rely on extensions of the SM which are motivated by other unsolved problems in particle physics. Examples of DM candidates include: weakly interacting massive particles (WIMPs), expected to arise in TeV extensions of the SM (in particular the LSP in supersymmetric models); the axion, introduced to solve the strong $CP$ problem; sterile neutrinos, hypothesised to explain the neutrino masses; mirror DM in a parity-symmetric universe. 

It is intriguing, however, that despite the possibly disparate origin of the visible and dark matter relic densities, and thus the different parameters determining their values, these densities differ only by a factor of a few. The apparent coincidence between the VM and DM is reinforced by the (as yet unverified) observations of the DM direct-detection experiments DAMA, CoGeNT and CRESST, 
which favour DM masses of a few GeV. This mass scale, which is very similar to that of the nucleons, may be considered together with the similar visible and dark total mass abundances to suggest that the number densities of the visible and dark relic species are related. It is then important to consider how such a relation can arise dynamically.

A tight connection between the VM and the DM densities is established dynamically if the visible and the dark sectors are both charged under a common particle number which remains \emph{always} conserved, and under which the two sectors develop compensating asymmetries. The common and always conserved symmetry has to be a generalisation of the ordinary baryon number, and we may thus refer to this paradigm as a ``baryon-symmetric universe''. 
The \emph{separation} of the baryonic and antibaryonic charge into the visible and the dark sectors, respectively, amounts to an asymmetry generated in an appropriately-defined particle number orthogonal to the generalised baryon number~\cite{Bell:2011tn}. We describe the symmetry structure and symmetry breaking pattern that gives rise to a baryon-symmetric universe in Sec.~\ref{sec:symm}. It should be already clear though, that the separation of baryons and antibaryons into two sectors requires non-equilibrium dynamics.

While establishing such a connection between the VM and DM densities is motivated phenomenologically, it is also desirable that this occurs within extensions of the SM which solve other (fundamental) particle-physics problems. Reference~\cite{Bell:2011tn} proposed that a baryon-symmetric universe can arise in supersymmetric models (SUSY) from Affleck-Dine (AD) dynamics~\cite{Dine:1995kz,Affleck:1984fy}. This mechanism was termed ``pangenesis'',
signifying the simultaneous generation of all matter. Examples of pangenesis were outlined in~\cite{Bell:2011tn} and subsequently in~\cite{Cheung:2011if}, although explicit dark sectors were not constructed.

In this paper, we discuss the mechanism of pangenesis in more depth, and construct specific models. We consider both gravity- or Planck-mediated SUSY breaking (PMSB) and gauge-mediated SUSY breaking (GMSB). In the latter case, we analyse the Affleck-Dine dynamics for arbitrary messenger mass, and present semi-analytical results which are of general relevance for AD baryogenesis in SUSY models, not just for those embodying pangenesis. We estimate the asymmetry generated, which in pangenesis quantifies the separation of baryonic and antibaryonic charge, and discuss the formation and decay of $Q$-balls, the non-topological solitons which can arise via the AD mechanism.  We produce plots which exhibit the feasibility of successful AD pangenesis (or baryogenesis). We also construct a class of simple dark sectors and describe the cascade of the dark baryonic charge down to the lightest dark sector particles. We discuss the nature of DM (which here can be `atomic'), and various pertinent constraints.

The possibility of a baryon-symmetric universe was first discussed by Dodelson and Widrow~\cite{Dodelson:1989ii,Dodelson:1989cq,Dodelson:1990ge}, and subsequently developed in various models which employed different dynamics for the separation of the baryonic from the antibaryonic charge. 
Besides the AD mechanism, these are: out-of-equilibrium decays of heavy particles~\cite{Kuzmin:1996he,Kitano:2004sv,Kitano:2005ge,Gu:2007cw,Gu:2009yy,An:2009vq,Davoudiasl:2010am,Gu:2010ft,Heckman:2011sw}, with such scenarios termed ``hylogenesis''~\cite{Davoudiasl:2010am}; the QCD phase transition~\cite{Oaknin:2003uv}; asymmetric freeze-out~\cite{Farrar:2005zd}; asymmetric freeze-in~\cite{Hall:2010jx}; a time-dependent background of a light scalar field~\cite{MarchRussell:2011fi}, and bubble nucleation due to a first-order phase transition~\cite{Petraki:2011mv}.

Related to the baryon-symmetric proposal, but distinctly different both in symmetry structure and observational signatures, is the scenario in which a baryonic asymmetry is originally generated in either the visible or the dark sector, and shared at a later time between the two sectors via chemical equilibrium. This possibility has been explored in Refs.~\cite{Nussinov:1985xr,Barr:1990ca,Barr:1991qn,Dodelson:1991iv,Kaplan:1991ah,Berezhiani:2000gw,Foot:2003jt,Foot:2004pq,Hooper:2004dc,Agashe:2004bm,Cosme:2005sb,Suematsu:2005kp,Suematsu:2005zc,Gudnason:2006ug,Gudnason:2006yj,Banks:2006xr,Dutta:2006pt,Berezhiani:2008zza,Khlopov:2008ty,Ryttov:2008xe,Foadi:2008qv,Kaplan:2009ag,Kribs:2009fy,Cohen:2009fz,Cai:2009ia,Frandsen:2009mi,Gu:2010yf,Dulaney:2010dj,Cohen:2010kn,Shelton:2010ta,Haba:2010bm,Buckley:2010ui,Chun:2010hz,Blennow:2010qp,Allahverdi:2010rh,Dutta:2010va,Falkowski:2011xh,Haba:2011uz,Chun:2011cc,Kang:2011wb,Graesser:2011wi,Frandsen:2011kt,Kaplan:2011yj,Cui:2011qe,Kumar:2011np,Graesser:2011vj,Oliveira:2011gn,Arina:2011cu,Kane:2011ih,Barr:2011cz,Lewis:2011zb,Cui:2011wk,D'Eramo:2011ec,Kang:2011ny}. Moreover, VM and DM may share the baryonic asymmetry in comparable amounts, without the need of a separate dark sector, if the baryonic asymmetry arises via the AD mechanism and is partially stored in the form of stable $Q$-balls comprising the DM of the universe~\cite{Kusenko:1997si}.
In this class of models, there is no symmetry, common to both the visible and the dark sector, which remains conserved throughout the cosmological history of the universe. However, both in this and in the baryon-symmetric scenario, the relic abundance of DM is due to a particle asymmetry, and both scenarios are thus often referred to as `asymmetric DM (ADM)' models.

This most generic property of baryon-symmetric models -- that there is an always conserved particle-number symmetry -- implies the possible existence of an associated gauge boson. The latter can be discovered in colliders and provides a channel for DM direct detection. This probe is absent in models with no unbroken baryonic symmetry. 

Finally, we note that it is possible to relate the visible and the dark matter relic abundances without relying on a dark asymmetry. 
The possibility that the AD mechanism is responsible for both the visible baryonic asymmetry and a symmetric DM abundance has been discussed in 
Refs.~\cite{Thomas:1995ze,Enqvist:1998en,Fujii:2002kr,Fujii:2002aj,Enqvist:2003gh,Dine:2003ax,Roszkowski:2006kw,McDonald:2006if,Kitano:2008tk,Shoemaker:2009kg,Higashi:2011qq,Doddato:2011fz,Mazumdar:2011zd,Kasuya:2011ix,Doddato:2011hx}. These models consider non-thermal DM production from  late $Q$-ball decays. Possible connections between the WIMP miracle and baryogenesis have been explored in Refs.~\cite{McDonald:2011zz,McDonald:2011sv,Cui:2011ab}.

In the next section we describe the symmetry structure and symmetry breaking pattern that give rise to a baryon-symmetric universe.
We discuss their realisation in supersymmetric models via the Affleck-Dine mechanism -- the process of pangenesis -- in Sec.~\ref{sec:pan}.
In Sec.~\ref{sec:AD}, we expound on the technical aspects of the AD dynamics. 
We construct a specific dark-sector model in Sec.~\ref{sec:models}, and discuss various cosmological constraints in Sec.~\ref{sec:DS constraints}. 
We discuss observational signatures of pangenesis in Sec.~\ref{sec:signatures}.

\section{How to build a baryon-symmetric universe}
\label{sec:symm}

\subsection{Symmetry structure}

The relic particle content of our universe is determined largely by the symmetries of the low-energy effective interactions, as described by the SM in the visible sector.

Most of the present VM mass density is carried by protons and heavier nuclei, whose stability is understood in terms of a 
global U(1) symmetry of the SM, generated by the baryon number operator of the visible sector, which we shall call $\Bv$. The charge $\Bv$ is additively conserved in perturbation theory, though violated by non-perturbative effects associated with anomalies.  These non-perturbative effects, which are very weak for temperatures below that of the electroweak phase transition, violate the global $\Bv$\footnote{Note that they respect a discrete subgroup of it, and thus do not endanger proton stability.}, but conserve a combination of the baryon and lepton symmetries of the SM, $\BLv$. 
The conservation of the global $\BLv$, and of $\Bv$ at low temperatures, ensures that an existing excess of baryons over antibaryons in the visible sector can be sustained, as 
indeed observed in today's low-energy universe.
Most ways of understanding the origin of the asymmetry between visible baryons and antibaryons require that the $\BLv$ symmetry was violated at some stage in the early universe, by as-yet unknown processes which subsequently became completely ineffective.\footnote{For completeness we note that the visible-sector baryon asymmetry can also be explained through electroweak baryogenesis, which uses the non-perturbative $\BLv$-conserving but $\Bv$-violating `sphaleron' processes mentioned earlier.}

If the DM relic abundance is also attributed to a particle-antiparticle asymmetry, then at least some of the dark-sector species have to be charged under a particle number which is additively conserved at low energies. We shall refer to this symmetry as the `dark baryon number' $\Bd$. 
In analogy to the visible sector, a dark baryonic asymmetry must have originated from processes which once violated $\Bd$ under the appropriate conditions for creation of a net $\Bd$ charge. Subsequently, these processes must have become unimportant, restoring $\Bd$ as a good symmetry of the low-energy interactions, and maintaining the existing dark baryonic asymmetry.

The visible and the dark baryonic asymmetries will be related if the processes which generated them violated the $\BLv$ and $\Bd$ symmetries while respecting a linear combination of the two. Quite generally, we may consider the two linear combinations\footnote{Under the definitions of \eq{eq:B-L,X def}, visible-sector fields are those with $B-L = X$, and dark-sector fields are those with $B-L = - X$. There may also be connector-type fields with linearly independent $B-L, \ X$.}
\beq
\bal{2}
B-L &\equiv \BLv - \Bd \\ 
X   &\equiv \BLv + \Bd \ .
\eal
\label{eq:B-L,X def}
\eeq
If the `generalised baryon number' $B-L$ remains conserved throughout the cosmological evolution of the universe, while $X$, $C$ and $CP$ violating interactions occur out of equilibrium, a net $X$ charge will be created, amounting to related $\BLv$ and $\Bd$ charge asymmetries,
\beq
\h \sqpare{\BLv} = \h \sqpare{\Bd} = \frac{ \h \sqpare{X} }{2} \ ,
\label{eq:asym rel}
\eeq 
where $\h[Q] \equiv \sum_i q_i [n(f_i) - n(\bar{f}_i)] / s$ is the charge-to-entropy ratio of charge $Q$.
The visible and the dark baryonic asymmetries compensate each other under the generalised baryon number $B-L$, and thus constitute a `baryon-symmetric universe'. The $X$ asymmetry quantifies the separation of baryonic and antibaryonic charge into the visible and dark sectors, respectively. The restoration of the $X$ symmetry in the low-energy late universe, together with the conservation of the $B-L$, ensure that the visible and the dark baryonic asymmetries are preserved separately.

The strict conservation of $B-L$ is guaranteed if it is a gauge symmetry. A gauged $B-L$ has multiple functionality for pangenesis. 
It precludes the possibility of the AD mechanism being operative along flat directions of the scalar potential with non-vanishing $B-L$, thus ensuring the validity of \eq{eq:asym rel}. It facilitates the decay of the heavier of the visible and dark sector LSPs, thus reducing a heavy relic that is unwanted in our scenario.
Moreover, the associated gauge boson offers an important probe into the dark sector, in both collider and DM direct-detection experiments.
We discuss these points in subsequent sections.

However, a gauged $B-L$ symmetry has to be broken to be consistent with the non-observation of a long-range force. We shall require that the breaking of the local $B-L$ leaves a global symmetry unbroken, under which all of the degrees of freedom appearing in the low-energy effective interactions transform according to their gauged $B-L$ charges. We discuss the spontaneous breaking of $B-L$ in the next subsection.

\subsection{Gauged and spontaneously broken $B-L$}
\label{sec:B-L break}

We wish to spontaneously break the gauged U(1)$_{B-L}$ symmetry without spoiling the symmetry structure discussed in the preceding subsection.  In particular, we must avoid introducing deleterious side-effects such as too-rapid proton and/or DM decay, and asymmetry erasure.  The first step to ensuring that these goals are met is to construct the gauged U(1)$_{B-L}$ as a diagonal subgroup of U(1)$_{\BLv} \times$U(1)$_{\Bd} \times$U(1)$_S$, where this last symmetry acts only to transform the Higgs field(s) responsible for the breaking of the gauged U(1)$_{B-L}$.  In other words, we seek to identify the gauged $B-L$ generator as
\beq
(B-L)_{\rm{gauged}} = \BLv - \Bd + S \equiv (B-L)_{\rm{global}} + S,
\eeq
with $S$, $\BLv$ and $\Bd$ [hence also $\BLv - \Bd \equiv (B-L)_{\rm{global}}$] being purely global symmetry generators when acting through any linear combination orthogonal to the one above.

We now explain this in more detail.  Suppose, for definiteness, that the visible sector is simply the minimal supersymmetric standard model (MSSM).  Then, as per the previous discussion, $\BLv$ is just the usual generator for global $B-L$ transformations of the quark and lepton superfields; in particular, it transforms neither the dark sector fields nor the Higgs sector responsible for breaking the gauged U(1)$_{B-L}$.  Similarly, $\Bd$ generates an analogous global phase symmetry in the dark sector, and does not act in either of the other two sectors.  The third generator $S$ transforms only the Higgs supermultiplets responsible for the breaking of the local U(1)$_{B-L}$; it does not transform any visible or dark sector fields.  A simple example of such a sector is exhibited through the superpotential terms
\beq
W \supset \beta \cal{T} ( \sigma \bar{\sigma} - v^2_{B-L} ),
\eeq
where $\sigma$ and $\bar{\sigma}$ constitute a vectorlike pair of chiral superfields, and $\cal{T}$ is trivial under all symmetries.  The U(1)$_S$ global symmetry transforms $\sigma$ and $\bar{\sigma}$ with equal and opposite phases, with $S$-charges $q_\sigma$ and $-q_\sigma$, respectively.  These superpotential terms contribute to the scalar potential the term
\beq
V \supset \beta^2 | \sigma \bar{\sigma} - v^2_{B-L } |^2
\eeq
which induces nonzero VEVs, $\langle \sigma \bar{\sigma} \rangle = v_{B-L}^2$, for the scalar components of the superfields $\sigma, \bar{\sigma}$.  We denote the scalar component of a superfield by the same symbol as the superfield, allowing context to resolve the ambiguity.

This VEV spontaneously breaks the global symmetry U(1)$_S$ and the gauge symmetry U(1)$_{(B-L)_{\rm{gauged}}}$ without breaking the global symmetries generated by $\BLv$ and $\Bd$.  The potential Nambu-Goldstone boson from spontaneous U(1)$_S$ breaking is eaten by the $B-L$ gauge field, so there is no massless scalar boson in the physical spectrum.  It is worth noting that requiring the MSSM sector to respect $(B-L)_{\rm{gauged}}$, and hence automatically $\BLv$, ensures that standard global baryon number and, separately, global lepton number emerge as accidental symmetries at the renormalisable level.

To ensure that U(1)$_S$ exists as an independent symmetry, we must choose the charge $q_\sigma$ judiciously.  For example, we should not choose $q_\sigma = 1$, for that would allow the superpotential terms $L H_u \sigma$ and $n^c \bar{\sigma}$ without violating overall $B-L$ conservation.\footnote{We denote the chiral superfields of the MSSM  by the standard notation $Q$, $L$, $u^c$, $d^c$, $e^c$ and $n^c$ for the quark doublet, lepton doublet, up antiquark, down antiquark, charged antilepton and neutral antilepton, respectively.} (The superfield $\sigma$ would then essentially contain a left-handed antineutrino.)  While we could impose independent U(1)$_S$ invariance even in this case, it is much more desirable to have this symmetry arise accidentally.  At the renormalisable level, this is ensured in the visible sector provided $\smash{q_\sigma \neq \pm 1/2,\pm1,\pm 2}$ so that neither $\sigma$ nor $\bar{\sigma}$ can couple to the monomials $LH_u$, $( n^c )^2$ and $n^c$.

We wish to extend this desirable feature to non-renormalisable effective operators as well.  The discussion below will focus on operators connecting the visible sector to the $\sigma$-sector, but an extension to dark sector fields will be required once that sector is specified.

Returning to the MSSM-gauge-invariant monomials $LH_u$ and $n^c$, we now need to consider higher order terms given generically by $LH_u \sigma^k$ and $LH_u \bar{\sigma}^k$ and similarly for $n^c$.  All such terms will be automatically absent if $q_\sigma \neq \pm 1/k$ for any positive integer $k$. 
At cubic order, the MSSM monomials $u^c d^c d^c$, $LLe^c$ and $Qd^c L$ all have $\BLv = -1$, so the same requirement forbids their couplings to positive-integer powers of $\sigma$ and $\bar{\sigma}$ also.  At higher order, in addition to other MSSM monomials carrying $\BLv = \pm 1$, there are monomials carrying $\BLv = \pm 2$ and $-3$~\cite{Gherghetta:1995dv}. Some of these monomials are listed in Table~\ref{table:monomials}. By multiplication, one can combine these `basis monomials' into a countable infinity of terms which must have $\BLv$ charges of the form $\pm n \pm 2m - 3p$, where $n$, $m$ and $p$ are non-negative integers.  The $\BLv$ charge of the general term $\mathcal{O}_{n,m,p} \sigma^k$ is $\pm n \pm 2m - 3p + q_\sigma k$, where $\mathcal{O}_{n,m,p}$ is any visible-sector monomial. For any rational charge $q_\sigma$, there exist monomials that can couple to a positive integer power of $\sigma$ or $\bar{\sigma}$ in a $(B-L)$-invariant way. Unless we allow the unorthodox possibility of making $q_\sigma$ irrational, it is thus not possible for U(1)$_S$ to be an accidental symmetry to all orders.\footnote{This conclusion has been reached assuming the visible sector is described by the MSSM.  If the visible-sector gauge symmetry is extended, then the form of the gauge-invariant effective operators will be more constrained.} Nevertheless, it is obvious that one can choose a rational $q_\sigma$ to forestall the explicit breaking of U(1)$_{\BLv} \times$U(1)$_S$ to U(1)$_{\BLv + S}$ to arbitrarily high order in terms of mass dimension.  Thus, we conclude that the spontaneous breaking of the gauged U(1)$_{B-L}$ poses no in-principle problem for proton decay and baryon asymmetry erasure.  By a straightforward extension, we will also be able to ensure sufficient DM stability and preservation of the dark-sector asymmetry.

Neutrino masses do not alter the above picture if neutrinos are Dirac particles. This includes the possibility of Dirac seesaw, which could explain the smallness of neutrino masses. If neutrinos are Majorana, then limits on the Majorana masses --and thus the $B-L$ breaking scale-- apply, in order to avoid wash-out of the visible and dark baryonic asymmetries. These limits depend on the seesaw scheme.

\section{Pangenesis via Affleck-Dine}
\label{sec:pan}

In the Affleck-Dine mechanism~\cite{Dine:1995kz,Affleck:1984fy}, the coherent oscillations of a scalar field generate a charge under a U(1) symmetry, when this symmetry is conserved by the low-energy couplings but explicitly broken by high-energy interactions. The formation of a condensate -- a homogeneous field configuration that possesses a time-dependant phase corresponding to a net U(1) charge -- relies on the scalar field acquiring a large vacuum expectation value (VEV) in the early universe. A large VEV amplifies small U(1)- and $CP$-violating terms, making them important at early times. If such terms are present, and if the scalar field possesses a large VEV but is not confined to a static configuration, i.e. it is free to spontaneously relax towards a lower energy state, the conditions for creation of an asymmetry~\cite{Sakharov:1967dj} are satisfied, and a U(1) charge is generated. Subsequently, the expansion of the universe causes the amplitude of the oscillating field to redshift, thus effectively restoring the U(1) symmetry and maintaining the generated charge. This charge may be transferred to lighter degrees of freedom transforming under the same symmetry if the condensate decays or thermalises via U(1)-conserving couplings.

The above dynamics can be naturally realised in supersymmetric theories with low-energy global U(1) symmetries. Because of the high degree of symmetry, the scalar potential of SUSY models generically possesses many flat directions (FDs) at the renormalizable limit.
The FDs are typically lifted by non-renormalizable operators, which may violate the U(1) symmetry that characterises the FD at low field values. 
The effective potential along a FD contains also SUSY-breaking contributions induced by the hidden sector, the thermal bath and the vacuum energy of the universe.  
The latter contribution dominates during inflation, and in the epoch of inflaton oscillations before reheating is completed, during which $H \gg \ms$, where $H$ is the Hubble parameter and $\ms$ stands for the soft mass scale induced by the SUSY-breaking hidden sector. 
The large vacuum energy of the universe and the absence of (unsuppressed) quartic terms along a FD can naturally drive the FD field to a large VEV at early times.
However, as $H$ decreases, a vanishing VEV becomes energetically more favourable. When the FD field begins to relax towards a zero VEV, 
the U(1) and $CP$ violating terms induce a time dependent phase for the field, which now thus carries a non-zero U(1) charge. 
If the U(1) symmetry of the FD is a symmetry of the full low-energy effective theory, the condensate decay and thermalisation processes will respect it, and transfer this charge to the lightest particles carrying the same conserved number. 

This is the case in the MSSM, where most of the FDs of the scalar potential carry baryon or lepton number. The AD baryo/lepto-genesis in the MSSM generates a $\BLv$ charge, which is carried initially by a squark or slepton condensate, and is subsequently transferred to the fermionic degrees of freedom by the $\BLv$-conserving interactions of the MSSM. The FDs of the MSSM have been listed in Ref.~\cite{Gherghetta:1995dv}.

Pangenesis can be realised in extensions of the MSSM along FDs with vanishing generalised $B-L$ charge and non-zero $X$ charge:
\begin{align}
D_{_{B-L}} &\equiv  \ \f^\dag  T_{_{B-L}} \f  =    0   \label{eq:D B-L}  \\
D_{_X}     &\equiv  \ \f^\dag  T_{_X}  \f     \ne  0   \label{eq:D X}  \ ,
\end{align}
where $T_{_{B-L}}$ and $T_{_X}$ are the $B-L$ and $X$ generators, respectively. The field $\f$ parametrises the FD, and is in general a linear combination of visible, dark and connector sector fields. 
A sum over these components of $\f$ is implied in Eqs.~\eqref{eq:D B-L} and \eqref{eq:D X}. The $X$ charge of the FD field is
\beq
q_{_{X,\f}} \equiv D_{_X} / |\f|^2 \ .
\label{eq:FD qX}
\eeq

Equation~\eqref{eq:D B-L} is warranted along FDs if the generalised $B-L$ is gauged. In this case, the corresponding $D$-terms in the scalar potential lift the directions which carry non-vanishing $B-L$, thus asserting that the AD mechanism cannot generate a net $B-L$ charge.\footnote{We note that, for field VEVs below the $(B-L)$-breaking scale $\MBL$, these $D$-terms are suppressed by a factor  $\smash{\pare{1+ \MBL^2 / \ms^2}^{-1}}$, where $\ms$ is the soft scale (see e.g.~\cite{Batra:2003nj}). To ensure that the AD field does not deviate from a path of vanishing $B-L$ during the important stages of its dynamics, we shall assume that $\MBL$ is smaller than the VEV $\fosc$ of the AD field at the onset of oscillations (respectively the VEV at $Q$-ball formation if applicable). 
Since the $(B-L)$-gauge boson can be as light as $500 \GeV$ for $\smash{\aBL \sim 10^{-2}}$ (see Sec.~\ref{sec:collider signatures}), this can always be satisfied. } 
While a gauged $B-L$ is most attractive for ensuring that it is always conserved, it is possible to justify a zero $B-L$ otherwise. In particular, a minimal K\"{a}hler potential for field directions with non-zero $B-L$ induces large positive mass-squared contribution along these directions during the vacuum-energy dominated era. This ensures that such field directions do not acquire large VEVs, and thus renders the AD mechanism inoperative along them.

Equation~\eqref{eq:D B-L} can be satisfied along field directions with either ${T_{_{B-L}} \f = 0}$, or ${T_{_{B-L}} \f \ne 0}$ and ${D_{_{B-L}} = 0}$. In the former case, the FD consists of connector-sector field(s) which carry both visible and dark baryonic charge, with $\BLv = \Bd \ne 0$. An example of this was presented in~\cite{Bell:2011tn}. 
The latter case corresponds to field directions along which $B-L$ is spontaneously broken while its expectation value remains zero. 
Such directions are necessarily made up by multiple fields, which may be purely visible and purely dark sector fields (carrying only visible or dark baryonic charge), or may also involve connector-sector fields with various charge assignments. Examples of this case were presented in~\cite{Cheung:2011if}.\footnote{We note though, that the two examples of Ref.~\cite{Cheung:2011if} are likely not viable:
They employ dimension-3 and dimension-4 monomials to lift the FDs and generate the asymmetry.
It is known, however, that dimension-4 monomials along a single flat direction do not produce enough asymmetry due to thermal corrections~\cite{Allahverdi:2000zd,Anisimov:2000wx,Anisimov:2001dp}. 
For a dimension-3 monomial, this problem can only become more severe. 
In addition, a dimension-3 monomial means that the direction considered is not $F$-flat already at the renormalisable level, and an unnaturally small coupling is required in order to generate sufficient asymmetry. This is an unnecessary complication arising from the specific choice of gauge-invariant monomial. Indeed, the scalar potential of supersymmetric theories typically possesses a large number of directions which are both $D$-flat and $F$-flat, at least at the renormalisable regime. This is in fact one of the basic reasons for which the Affleck-Dine mechanism is implemented in supersymmetric models only: that the quartic couplings in the scalar potential vanish identically along flat directions, at the supersymmetric regime, due to the symmetries of the theory.}

In the next section, we present explicitly the FD potential which gives rise to the AD dynamics, in the PMSB and the GMSB scenarios.
We present extensions of the MSSM in which these dynamics operate under the conditions of \eqs{eq:D B-L} and \eqref{eq:D X} in Sec.~\ref{sec:models}.

\section{The Affleck-Dine Dynamics}
\label{sec:AD}

The directions of the scalar potential which are flat in the renormalisable and supersymmetric regime are typically lifted by non-renormalisable and SUSY-breaking interactions.

We consider non-renormalizable terms in the superpotential of the type
\beq
W_\rm{nr} = \frac{ \F^d }{d \, \Mnr^{d-3} } \ , \quad d \geqslant 5 \ ,
\label{eq:Wnr}
\eeq
which lift a FD parameterised by the scalar degree of freedom $\F$. It will often be convenient to reparametrise the scalar component, denoted also by $\F$, in terms of amplitude and phase, as per
\beq
\F \equiv \frac{\f}{\sqrt{2}} e^{i \th}\ .
\label{eq:fth}
\eeq
The dimension $d$ of the operator in \eq{eq:Wnr} is determined by the particle content and the charge assignments of the model under consideration. This term violates explicitly the U(1) symmetry characterising the FD at low field values. For successful pangenesis, this symmetry has to be U(1)$_X$.
The scale $\Mnr$ at which U(1)$_X$ is violated may be as high as the Planck scale, and as low as the requirement of non-erasure of the baryonic asymmetries allows.\footnote{Note that this scale can be effectively larger than the Planck scale when the operator arises within a Froggatt-Nielsen framework (see e.g.~\cite{Fujii:2001zr,Harnik:2004yp}).} We examine this issue in Sec.~\ref{sec:models}. The superpotential coupling of \eq{eq:Wnr} contributes to the potential along the FD the term
\beq
V_\rm{SUSY} = \abs{ \frac{\F^{d-1} }{\Mnr^{d-3} }}^2 \ .
\label{eq:VSUSY}
\eeq
Despite arising from an $X$-violating coupling in the superpotential, this term respects the U(1) symmetry of the FD. It dominates and stabilises the potential at large field values.

SUSY breaking also contributes to the potential along FDs. There are three sources which are relevant for the AD dynamics: the vacuum energy of the universe, the thermal bath and the hidden sector which dominates SUSY breaking in the late universe. 

During inflation and in the epoch of inflaton oscillations, the universe is dominated by the energy stored in the inflaton field. 
The coupling of the inflaton to the AD field can induce a large effective mass for the latter. The mass term depends on the K\"{a}hler potential: while a minimal coupling gives rise to a positive mass-squared term, non-renormalisable couplings in the K\"{a}hler potential can generate a negative mass-squared contribution for the AD field, 
\beq
V_H = -c H^2 |\F|^2  \ ,
\label{eq:H mass}
\eeq
where $c \sim 1$. A detailed study of the Hubble-induced terms for various types of the K\"{a}hler potential can be found in~\cite{Kasuya:2006wf}. 
A Hubble-induced negative mass-squared contribution ensures that the FD field acquires a large VEV at early times, which is essential for the efficacy of the AD mechanism. 
The VEV of the field is determined by the balance between the terms of \eqs{eq:VSUSY} and \eqref{eq:H mass}:
\beq
\frac{\f_\rm{min}}{\sqrt{2}} \approx  \Mnr  \pare{\frac{H}{\Mnr}}^\frac{1}{d-2} \ .
\label{eq:fmin}
\eeq

The SUSY breaking in the hidden sector also induces a soft mass term along the FD, whose precise form depends on the way SUSY breaking is communicated.
The mass-squared contribution from the hidden sector has to be positive at low energies (small VEVs) to ensure the restoration of the symmetry
\beq
V_s = \ms^2 |\F|^2 + ... \ ,
\label{eq:Vs}
\eeq 
where the dots stand for corrections which depend on the SUSY-breaking mediation mechanism, and which become important with increasing field VEV. In the following sections, we consider both the PMSB and the GMSB scenarios, and specify the form of the term of \eq{eq:Vs}. 

In addition, the dilute thermal bath of the universe during the epoch of inflaton oscillations generates a thermal potential for the AD field~\cite{Allahverdi:2000zd,Anisimov:2000wx}. 
The details of thermalisation of the inflaton decay products depend largely on the model of inflation, and can be affected by large VEVs developed along flat directions of the scalar potential~\cite{Allahverdi:2005mz,Allahverdi:2007zz}. Here we assume that the inflaton couples generically to both the visible and the dark sector, and model the temperature of the inflaton decay products before the completion of reheating by~\cite{Turner:1983he}
\beq
T \approx \pare{H T_R^2 \Mp }^{1/4} \ ,
\label{eq:T>TR}
\eeq
where $T_R$ is the reheating temperature. This gives rise to two contributions:
(i) The relativistic degrees of freedom which couple directly to the AD field induce a thermal mass term for the latter:
\beq
V_T \supset c_k y_k^2 \: T^2 \: |\F|^2 \ , \quad \rm{for} \quad y_k \f < T \ .  
\label{eq:VT<}
\eeq
Here $c_k$ are constants of order 1, and $y_k$ stand for the Yukawa or gauge coupling constants of the AD field to these particles, defined such that $y_k \f$ is the effective mass they pick up when the AD field has a non-zero VEV $\f$. 
A summation over all species for which $y_k \f < T$ is implied in \eq{eq:VT<}. 
The degrees of freedom for which this condition does not hold, obtain a large mass and become non-relativistic. The suppression of their contribution to the thermal mass of the AD field at $y_k \f / T >1$ is exponential.
(ii) If some of the heavy degrees of freedom for which $y_l \f > T$, also couple to other relativistic particles, they can be integrated out, and generate effective couplings of the AD field to the thermal bath. The large $\f$-dependant mass of these mediators affects the running of their couplings to the light particles, and results in a potential which is logarithmic in the amplitude of the AD field~\cite{Anisimov:2000wx}  
\beq
V_T \supset \tilde{c}_l \: \frac{y_l^4}{16 \p^2} \: T^4 \ln \pare{\frac{y_l^2 |\F|^2}{T^2}} \ , \quad \rm{for} \quad y_l \f > T \ ,  
\label{eq:VT>}
\eeq
where again $\tilde{c}_l = \mathcal{O}(1)$ and summation over all degrees of freedom for which $y_l \f > T$ is implied. The $1/16 \p^2$-factor reflects the loop suppression of this contribution.\footnote{We do not use the interpolation between \eqs{eq:VT<} and \eqref{eq:VT>} which was introduced in Ref.~\cite{Kasuya:2001hg}, ${V_T \sim T^4 \ln \pare{1+y^2 |\F|^2 / T^2}}$, since it overestimates the contribution of \eq{eq:VT>} to the effective mass-squared of the condensate by $16 \p^2 / y_l^4$. This factor is significant even for very large couplings, $y_l \sim 1$. This fact is important in determining the onset of oscillations of the AD field and thus the generated asymmetry (see \eqs{eq:Hosc} and \eqref{eq:eta}).}

The effective potential felt by the AD field at low energies is
\beq
V_\rm{eff} = V_s + V_T \ .
\label{eq:Veff}
\eeq
The interplay between the hidden-sector/thermal and the Hubble-induced mass terms determines the spontaneous relaxation of the AD field towards its late-time ground state, as governed by its equation of motion $\ddot{\F} + 3 H \dot{\F} = -\partial V/\partial \F^*$. When $H$ becomes sufficiently small, the hidden-sector/thermal mass term drives the evolution of the field because $V_\rm{eff}'(\f) > V_H'(\f)$, and the field starts oscillating around the origin. This occurs at
\beq
\Hosc \approx \w \equiv  \pare{ \frac{V_\rm{eff}'(\f)}{\f} }^{1/2} \ .
\label{eq:Hosc}
\eeq
The amplitude of the field at the onset of oscillations, $\fosc$, is given by \eq{eq:fmin} with ${H = \Hosc}$.
It is during this \emph{non-equilibrium} process of relaxation towards the lowest energy state that the asymmetry is generated.

For this to occur, it is of course necessary that U(1)-breaking terms operate along the FD. Such terms are generated as a result of SUSY breaking\footnote{In AD scenarios in which multiple flat directions are excited, U(1) violation arises also in the terms derived directly from the superpotential. This enhances the asymmetry produced~\cite{Senami:2002kn,Enqvist:2003pb,Kamada:2008sv}. This feature was employed in the model of pangenesis constructed in~\cite{Bell:2011tn}.} and the U(1)-violating superpotential interaction of \eq{eq:Wnr}:
\beq
V_A =  A \frac{ \F^d }{d \, \Mnr^{d-3} }  +  \hc   
\label{eq:VA}
\eeq
In the above, the parameter $A$ is again determined by the mediation mechanism of SUSY breaking to the FD fields.
There may also be a Hubble-induced contribution, $\delta A \sim H$. However, such a contribution is absent in $F$-term inflation models, in which the interaction leading to it is proportional to the inflaton field and thus averages to zero in the epoch of inflaton oscillations~\cite{Kasuya:2008xp}. We shall omit this contribution in our analysis.

\medskip

We ultimately want to estimate the $X$-charge density generated in the condensate via the AD mechanism, 
\beq
\nX = i \qX \pare{\dot{\F}^* \F - \F^* \dot{\F} } \ ,
\eeq
where $\qX$ is the $X$ charge of the FD field. The equations of motion in an FRW background give
\beq
\dot{n}_{_X} + 3 H \nX = i \qX \pare{ \frac{d V}{d \F^*} \F^* - \frac{d V}{d \F} \F } \ . 
\label{eq:dot nX}
\eeq
Only the $X$-violating terms of the FD potential contribute to the right-hand side of \eq{eq:dot nX}. In the setup we consider here, $X$ violation arises only from the $A$-terms of \eq{eq:VA}, thus \eq{eq:dot nX} becomes
\beq
\dot{n}_{_X} + 3 H \nX =  2\qX \frac{|A| \pare{\f/\sqrt{2}}^d }{ \Mnr^{d-3}} \sin \pare{ \th_A + d \cdot \th } \ ,
\label{eq:dot nX expl}
\eeq
where we parametrise $A = |A| \exp (i \th_A)$, and $\f$ and $\th$ are defined in \eq{eq:fth}.
The above equation reflects two essential features of the AD mechanism: 
(i) the large VEV of the AD field amplifies the effect of the otherwise suppressed U(1)-violating interactions;
(ii) the $CP$ violation necessary for the generation of an asymmetry arises from a misalignment between the phase of an initial field configuration and the phase of the U(1)-violating terms.

The $CP$ violation is the final element we need to identify in this set-up. At early times, the field is dynamically confined at a large VEV by the Hubble-induced negative mass term, which generates a curvature of order $H$ along the radial direction, around the minimum. The $A$-term generates $d$ minima and maxima along the angular direction in the internal phase space of the field, with radial curvature of order $A$. If $A \ll H$, the angular curvature of the potential is much smaller that the radial curvature, and the field picks a random initial phase.\footnote{If A does not contain a Hubble-induced contribution, this is typically the situation. However, if the hidden-sector potential is very flat, oscillations may begin very late, and $A\sim \Hosc$ can be realised, even in the absence of Hubble-induced $A$-terms (see Sec.~\ref{sec:GMSB}).} 
When the radial oscillations begin, the mismatch between this initial phase and the phase of the $A$-term kicks the field in the angular direction, thus generating a time-dependant phase and a net charge for the condensate.

The charge is thus generated at the onset of oscillations, within a Hubble time, while $CP$ violation is effective. It can be estimated from \eq{eq:dot nX expl}, by ignoring the instantaneous expansion of the universe: $\delta \nX \approx \dot{n}_{_X} \delta t$ with $\delta t \sim H^{-1}$, which gives
\beq
\nX (t_\rm{osc}) 
\ \approx \ 2\qX \sin \delta \frac{|A| \: \pare{\fosc / \sqrt{2}}^d}{\Mnr^{d-3} \Hosc}
\ \approx \ \qX \sin \delta  \: |A|  \fosc^2 \ ,
\label{eq:nX tosc}
\eeq
where in the second equality we used \eq{eq:fmin}. 
The phase $\delta$ represents the effective $CP$ violation, and we shall typically assume $\sin \delta \sim 1$.

The asymmetry generation occurs in the epoch of inflaton oscillations, during which there is rapid entropy production. The $X$-charge-to-entropy ratio should then be calculated at reheating, after which the entropy per comoving volume remains constant. The $X$-charge density decreases as $\nX \propto a ^{-3}$. During inflaton oscillations, the Hubble rate scales like ${H \propto a^{-3/2}}$. This remains valid even after the inflaton has decayed if most of the energy is stored in non-relativistic particles. This can happen if large VEVs of scalar fields after inflation induce large masses for the inflaton decay products~\cite{Allahverdi:2005mz,Allahverdi:2007zz}. As these VEVs redshift, various degrees of freedom become light, and contribute to the relativistic energy density of the universe and to the various thermalisation processes. 
Assuming that the transition from a matter-dominated-like universe to the radiation-dominated era signals the complete thermalisation of all degrees of freedom (reheating), the $X$-charge density at reheating is estimated to be\footnote{There may be a period of radiation domination between this transition (at Hubble rate $H_\rm{m-r}$) and complete thermalisation (at $H_R$). For $\Hosc > H_\rm{m-r} > H_R$, \eq{eq:nX t} should in this case be replaced by $\nX(t_R) / \nX(t_\rm{osc})=  \sqpare{H_\rm{m-r} / \Hosc}^2 \sqpare{H_R / H_\rm{m-r}}^{3/2} =  \sqpare{H_R / \Hosc}^2 \sqpare{H_\rm{m-r} / H_R}^{1/2}$. This means that the $X$-charge-to-entropy ratio becomes enhanced by a factor of 
$\sqpare{H_\rm{m-r} / H_R}^{1/2}$ compared to \eq{eq:eta}. Considering the various upper bounds on the reheating temperature (determined by $H_R$) from cosmological considerations, this opens additional parameter space for successful asymmetry generation via the AD mechanism. In the rest of the analysis, we shall only focus on the ${H_\rm{m-r} = H_R}$ case, as described in the text.}
\beq
\nX(t_R) = \nX(t_\rm{osc}) \frac{H_R^2}{\Hosc^2} \ .
\label{eq:nX t}
\eeq
The entropy density at reheating is $s(t_R) = 4 H_R^2 \Mp^2 /T_R$, thus the $X$-charge-to-entropy ratio is
\beq
\hX = \frac{\nX(t_R)}{s(t_R)} =  \frac{\nX(t_\rm{osc}) T_R}{4 \Hosc^2 \Mp^2} \ .
\label{eq:eta gen}
\eeq
Using the expression on the right-hand side of \eq{eq:nX tosc}, this gives
\beq
\hX 
\ \approx \  \frac{\qX \sin \delta}{4} \: \frac{T_R}{\Mp^2} \frac{|A| \fosc^2}{ \Hosc^2}  
\ \approx \  \frac{\qX \sin \delta}{2} \: \frac{|A| T_R}{\Mp^2} \pare{ \frac{\Mnr}{\Hosc} }^\frac{2(d-3)}{d-2} \ ,
\label{eq:eta}
\eeq
where in the second equality we made use of \eq{eq:fmin}.
We will use both forms of \eq{eq:eta} in our analysis below.

An important quantity for the late-time evolution of the condensate is the ellipticity of the field orbit in its internal phase space:
\beq
\e \equiv \frac{\dot{\th}}{\w } = \frac{\nX/\qX}{\w \:  \f^2} = \sin \delta \frac{|A|}{\Hosc}\ .
\label{eq:ellip}
\eeq
The quantity $\w$, defined in \eq{eq:Hosc}, is the maximum rotational velocity that the AD field can develop at the onset of oscillations, $\dot{\th} \leqslant \w$. Maximum $\dot{\th}$ occurs if the $A$-terms which provide the U(1) violation are dominant, $A=\Hosc$, and the $CP$-violation is maximal, $\sin \delta \simeq 1$.

\subsection{Gravity-mediated SUSY breaking: asymmetry generation}
\label{sec:PMSB}

We shall now specify the form of \eq{eq:Vs} in the PMSB scenario, combine all the contributions to the FD potential discussed above, estimate the asymmetry generated, and identify the parameter space for successful pangenesis.

The potential along a FD in the PMSB scenario can be described by~\cite{Enqvist:1997si}
\beq
V_\rm{AD} =  \ms^2 |\F|^2 \sqpare{1 + K \ln \pare{ \frac{|\F|^2}{\Mp^2} } } + V_T  -c H^2|\F|^2  
+ \pare{ A \frac{ \F^d }{d \: \Mnr^{d-3} }  +  \hc  } + \abs{ \frac{\F^{d-1} }{ \Mnr^{d-3} }}^2 \ ,
\label{eq:V PMSB}
\eeq
where 
\beq
A \approx \ms \ ,
\label{eq:A PMSB}
\eeq
and the soft scale is of order the gravitino mass in PMSB, $\ms \sim \mg$. The first term in \eq{eq:V PMSB} is the hidden-sector SUSY breaking mass term. The correction proportional to $K$  arises radiatively at 1-loop, and depends on the FD~\cite{Enqvist:1997si,Enqvist:2000gq}. Typically $|K| \sim 0.01 - 1$. The sign of $K$ can be important for the late-time evolution of the AD condensate, as we discuss in Sec.~\ref{sec:Q}.  The thermal potential $V_T$ is given by \eqs{eq:VT<} and \eqref{eq:VT>}. 

\begin{figure}[t]
\centering
\includegraphics[width=8.3cm]{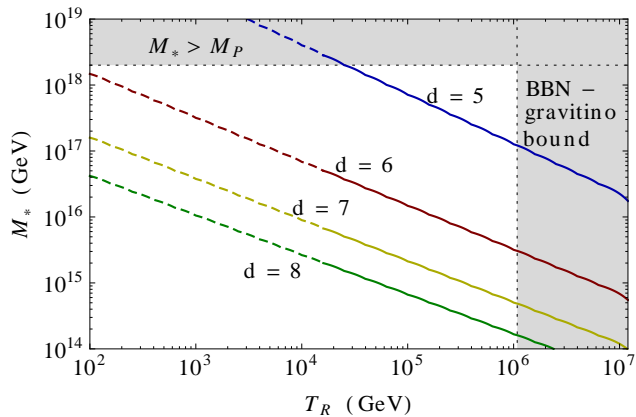}
\caption{The $T_R - M_*$ values which yield the correct asymmetry in PMSB. The different curves correspond to various values of the dimension $d$ of the monomial which lifts the flat direction. From top to bottom: $d=5$ (blue), $d=6$ (red), $d=7$ (yellow), $d=8$ (green). The dashing of the lines denotes the excluded region from the requirement that $Q$-balls decay before the LSP freezes out, $T_Q \gtrsim 10 \text{ GeV}$, which sets $T_R \gtrsim 6 \cdot 10^3 \text{ GeV}$ (see \eq{eq:TR limit PMSB}). This bound does not apply if $Q$-balls do not form, as is the case for flat directions with $K>0$, and even for $K<0$ it may be relaxed (see text for discussion). The vertical dotted line at $T_R = 10^6 \text{ GeV}$ denotes the upper limit on the reheating temperature in PMSB scenarios, from considering the production and decay of gravitinos~\cite{Kawasaki:2008qe}. The horizontal dotted line is drawn at the reduced Planck mass. 
For this plot, we used $\eta_{_X} = 10^{-9}, \ q_{_X}=1, \ \sin \delta = 1, \ y=10^{-2}, \ \ms = 500 \text{ GeV}$ and $K = -0.01$.}
\label{fig:PMSB TR vs Mnr}
\end{figure}

The AD field starts oscillating around the origin of the potential when, according to \eq{eq:Hosc}, the Hubble expansion rate becomes
\beq
\Hosc^2 \approx  \ms^2 
+ c_k y_k^2 \: \Tosc^2 
+ \frac{\tilde{c}_l y_l^4 \: \Tosc^4}{16 \p^2 \f^2} \ , 
\label{eq:Hosc PMSB}
\eeq
where summation over all degrees of freedom for which $y_k \f <T$ and $y_l \f > T$ is implied.
Equations~\eqref{eq:fmin}, \eqref{eq:T>TR} and \eqref{eq:Hosc PMSB} have to be solved together to obtain the time at which oscillations start.  
Typically, the thermal contributions to the potential force oscillations to start earlier, and thus suppress the asymmetry generation, as per \eq{eq:eta}. This, in turn, means that a larger reheating temperature and/or a larger $\Mnr$ are required in order to obtain a sufficiently large $X$-charge-to-entropy ratio.
The precise effect of the thermal corrections on the onset of oscillations and the asymmetry generation depends non-trivially on $\Mnr, T_R$ and the various coupling constants $y_k$, and more detailed analyses can be found in Refs.~\cite{Asaka:2000nb,Fujii:2001zr}. 
Here we intend to only roughly sketch the parameter space which produces the correct asymmetry. We thus pick a sample value for a coupling, $y \sim 10^{-2}$, calculate the onset of oscillations numerically, and present the combination of $\Mnr, T_R$ values which yield $\hX \simeq 10^{-9}$ in Fig.~\ref{fig:PMSB TR vs Mnr}.\footnote{For the purpose of illustration, in all of the graphs we use $\ms=500 \GeV$. The equations of course retain their explicit dependence on the soft scale. While LHC has already placed stringent bounds on the masses of the superpartners, it is still possible for some of the sparticles to be at or below 500 GeV~\cite{Kats:2011qh,Papucci:2011wy}. In addition, pangenesis occurs along flat directions which involve both MSSM and dark-sector scalars. The masses of the latter are not constrained, and may be lower than in the visible sector. A lower soft scale in the dark sector may in fact be desirable: it can be generated dynamically, and may be responsible for the GeV-scale DM mass (see discussion in Sec.~\ref{dsGMSB}). The soft scale relevant for mixed visible-dark flat directions is an average of the visible and the dark soft scales, and can be thus comfortably below 500 GeV. }

The reheating temperature is constrained from the production and decay of gravitinos in the early universe. Gravitinos are produced from scatterings of thermal particles in the primordial plasma. Even if they are unstable, which is typically the case in PMSB scenarios, their lifetime is extremely long: a gravitino lighter than 10~TeV decays at time scales larger than 1~sec, after BBN has started. The successful predictions of BBN can then be retained only if the abundance of gravitinos is sufficiently small. This yields an upper limit on the reheating temperature. A recent detailed analysis was presented in Ref.~\cite{Kawasaki:2008qe} which finds that for an unstable gravitino of about 500~GeV mass, the reheating temperature has to satisfy ${T_R \lesssim 10^6 \GeV}$.

Given this constraint, \eqs{eq:eta} and \eqref{eq:Hosc PMSB} imply that for $d=4$ the asymmetry generated is not sufficient, even for $\Mnr$ as high as $\Mp$ (this is true even in the limit $y \to 0$). However, for $d \geqslant 5$, it is possible to get $\hX \approx 10^{-9}$ for $\Mnr \lesssim \Mp$ while respecting the upper limit on the reheating temperature, as shown in Fig.~\ref{fig:PMSB TR vs Mnr}. 

\subsection{Gauge-mediated SUSY breaking: asymmetry generation}
\label{sec:GMSB}

\subsubsection{Effective potential and onset of oscillations}
\label{sec:GMSB--onset}

We now specify the hidden-sector SUSY-breaking mass term of \eq{eq:Vs}, in the GMSB scenario, present the complete potential, and follow the introductory analysis in order to estimate the asymmetry generation and investigate the parameter space.

The potential along a flat direction due to GMSB from the hidden sector has been calculated in the limit of small and large field VEVs in Ref.~\cite{deGouvea:1997tn}. 
Here, we adopt the interpolation introduced in Ref.~\cite{Doddato:2011fz}. Including the terms already discussed, we consider the following potential
\begin{multline}
V_\rm{AD} =  \ms^2 \Mm^2 \ln^2 \pare{1+ \frac{|\F|}{\Mm}} \sqpare{1 + K \ln \pare{ \frac{|\F|^2}{\Mm^2} } } + V_T   \\ 
-c H^2|\F|^2  + \pare{ A \frac{ \F^d }{d \, \Mnr^{d-3} }  +  \hc  } + \abs{ \frac{\F^{d-1} }{ \Mnr^{d-3} }}^2 \ ,
\label{eq:V GMSB}
\end{multline}
where $\ms$ is the soft mass scale and $\Mm$ is the messenger scale.\footnote{In \eq{eq:V GMSB}, we have omitted the contribution from PMSB which becomes important only at large field amplitudes, when $\smash{\mg^2 \f^2 \gtrsim \ms^2 \Mm^2}$. Using \eq{eq:Mm}, we find that this occurs at ${\f \gtrsim 10^{-3} \Mp \simeq 10^{15} \GeV}$, \emph{independently} of both the gravitino mass and the soft scale. However, the analysis of Sec.~\ref{sec:GMSB param} shows that the oscillations of the AD field in the GMSB scenario always start at lower field VEVs. It is thus justified to omit the PMSB contribution in the subsequent analysis.} 

The first term in \eq{eq:V GMSB} resembles a quadratic potential, $\ms^2 \f^2 /2$, in the $\f \ll \Mm$ limit, and it flattens out for $\f \gtrsim \Mm$. This is because in this regime, the gauge fields acquire large masses $\sim g \f > \Mm$ and the transmission of SUSY breaking is suppressed. 
As in the PMSB scenario, the correction proportional to $K$ arises from 1-loop corrections, and depends on the FD.

The $A$-terms in GMSB are generated at 2-loop and are thus suppressed with respect to the soft scale. Following~\cite{Doddato:2011fz}, we model the parameter $A$ in Eq.~\eqref{eq:VA} by
\beq
A = \mg + \frac{a_o  \ms}{\pare{1+\frac{|\F|^2}{\Mm^2}}^{1/2}} \ .
\label{eq:A GMSB}
\eeq
Here the first term is the contribution from PMSB, and the second term includes the loop-suppression factor, $a_o \sim 0.01$, and takes into account that GMSB is suppressed at large field VEVs $\f \gg \Mm$.

\medskip

For the potential of \eq{eq:V GMSB}, the oscillations around the origin begin according to \eq{eq:Hosc} at
\beq
\pare{ \frac{\Hosc}{\ms} }^2 = \frac{\ln \pare{ 1 + \frac{\fosc}{\sqrt{2}\Mm} }}{ \frac{\fosc}{\sqrt{2} \Mm} \pare{ 1 + \frac{\fosc}{\sqrt{2} \Mm} } }  \ ,
\label{eq:Hosc GMSB}
\eeq
where we omitted corrections which are proportional to $K$ and logarithmic in $\fosc/\sqrt{2}\Mm$, and also the thermal corrections. The latter can be unimportant in the regimes allowed by other constraints for reasonable values of the FD couplings to other fields. We discuss this further in Sec.~\ref{sec:GMSB param}.
At the onset of oscillations the field VEV is determined by the minimum of the potential prior to oscillations, \eq{eq:fmin}, which for later convenience we rewrite as
\beq
\frac{\fosc}{\sqrt{2} \Mm} \simeq b \pare{ \frac{\Hosc }{\ms } }^{ \frac{1}{d-2} } \ , 
\label{eq:osc init}
\eeq
where we defined
\beq
b \equiv \pare{ \frac{\ms \Mnr^{d-3}}{\Mm^{d-2}} }^{ \frac{1}{d-2} } \ .
\label{eq:b}
\eeq

Equations \eqref{eq:Hosc GMSB} and \eqref{eq:osc init} yield the amplitude of the field, $\fosc$, and the Hubble parameter, $\Hosc$, when oscillations begin. The dimensionless parameter $b$ defines different parametric regimes, which we list in Sec.~\ref{sec:GMSB param}. In each case, we estimate the generated asymmetry from \eq{eq:eta}. The asymmetry depends on the reheating temperature $T_R$, which in the GMSB scenario is constrained from the requirement that gravitinos do not overclose the universe. We now turn to this issue.

\subsubsection{Constraints from gravitinos}
\label{sec:GMSB--grav}

The gravitino couples to the other species with strength inversely proportional to its mass. In GMSB, gravitinos are expected to be light and stable, thus scatterings and decays of thermalised supersymmetric particles in the primordial plasma can produce a significant gravitino relic abundance~\cite{Moroi:1993mb,deGouvea:1997tn}.

For $\mg \gtrsim 100 \keV$, scattering processes produce gravitinos more efficiently than decays, and result in a gravitino relic density
\beq
\W_{3/2} \approx 0.2 
\pare{ \frac{1 \GeV}{\mg} }  
\pare{ \frac{m_\rm{gl}}{1 \TeV} }^2
\pare{ \frac{T_R}{10^{8} \GeV} } \ ,
\label{eq:W3/2}
\eeq
where $m_\rm{gl}$ is the gluino mass. We shall require that gravitinos are a subdominant contribution to the dark matter of the universe, i.e. $\W_{3/2} \lesssim 0.1 \, \W_\rm{DM}$. This yields an upper limit on $T_R$, which becomes less severe if the gravitino is heavy.

For smaller masses, $100 \keV \gtrsim \mg \gtrsim 1 \keV$, decays of supersymmetric particles produce gravitinos very efficiently. It is then necessary to ensure that squarks and sleptons are never very abundant, which implies that the reheating temperature has to be lower than their mass. To be conservative, we will require $T_R \lesssim 100 \GeV$ in this regime. 

For even lower masses, $1 \keV \gtrsim \mg \gtrsim 100 \eV$, if the reheating temperature is too high, the gravitinos thermalise, decouple while relativistic, and overclose the universe. We therefore require again $T_R \lesssim 100 \GeV$ to ensure that the relevant scattering processes are never in (relativistic) equilibrium, and gravitinos do not thermalise despite their large coupling to the MSSM particles. If $\mg < 100 \eV$, the gravitino relic abundance after their freeze-out is subdominant, and there is no limit on the reheating temperature.

While a larger gravitino mass reduces the gravitino relic abundance (which is unwanted for our scenario), thus allowing a higher reheating temperature without overproducing gravi\-tinos, it also decreases the decay rate of the NLSP, since the gravitino couples more weakly. If the NLSP was thermalized in the early universe and the gravitino is too heavy, the late decay of the NLSP freeze-out abundance can spoil BBN. We therefore again require ${T_R \lesssim 100 \GeV}$ in this regime. The value $\tilde{m}$ of the gravitino mass, for which this bound becomes relevant, depends on the nature and the mass of the NLSP. It is more stringent for bino than for stau, and quite relaxed for sneutrino NLSP~\cite{Kawasaki:2008qe}:
\beq
 \tilde{m} \equiv \left\{
\bal{2}
  1 \GeV,& \quad \rm{for \ NLSP} = \tilde{b}  \\
 10 \GeV,& \quad \rm{for \ NLSP} = \tilde{\t} \\
100 \GeV,& \quad \rm{for \ NLSP} = \tilde{\n} \ .
\eal
\right.
\label{eq:m3/2 con}
\eeq

In summary, we consider the following constraints on the reheating temperature:
\beq
T_R \lesssim \left\{
\bal{5}
&100 \GeV,          \quad     &&\rm{for} \quad    \ \mg \  \gtrsim \tilde{m}  \\
&4 \cdot 10^7 \GeV \pare{\frac{\mg}{1 \GeV}} \pare{\frac{500 \GeV}{\ms}}^2,  \quad  &&\rm{for} \quad 100 \keV \lesssim  \ \mg \  \lesssim \tilde{m}  \\
&100 \GeV,          \quad     &&\rm{for} \quad 100 \eV  \ \lesssim \mg \  \lesssim 100 \keV  \\
&\rm{no \ limit},   \quad     &&\rm{for} \quad                   \mg \  \lesssim 100 \eV \ ,  
\eal
\right.
\label{eq:TR limit}
\eeq
where the constraint of the first line arises from BBN considerations, while the rest ensure underabundant gravitino DM.

Finally, we note that the LSP, instead of being the gravitino, could also reside in the dark sector. We outline a model where this is possible in Sec.~\ref{dsGMSB}. The gravitino then decays to this lighter particle. This would ease the bound on the reheating temperature from the requirement that the LSP is a subdominant contribution to the dark matter. In addition, the NLSPs in the visible sector decay to the dark sector via the $(B-L)$-gaugino, resulting typically in a shorter NLSP lifetime than in the case of gravitino LSP. This would ease constraints from NLSP decays during BBN. A more detailed study of the constraints on the reheating temperature in this case is, however, beyond the scope of this work.

\subsubsection{Parameter space analysis}
\label{sec:GMSB param}

In what follows we estimate the charge-to-entropy ratio generated via the AD mechanism in the GMSB scenario, according to what was described in Sec.~\ref{sec:GMSB--onset}. We apply the constraints discussed in Sec.~\ref{sec:GMSB--grav}, and identify the available parameter space.

\begin{figure}[t]
\centering
\includegraphics[width=7.2cm]{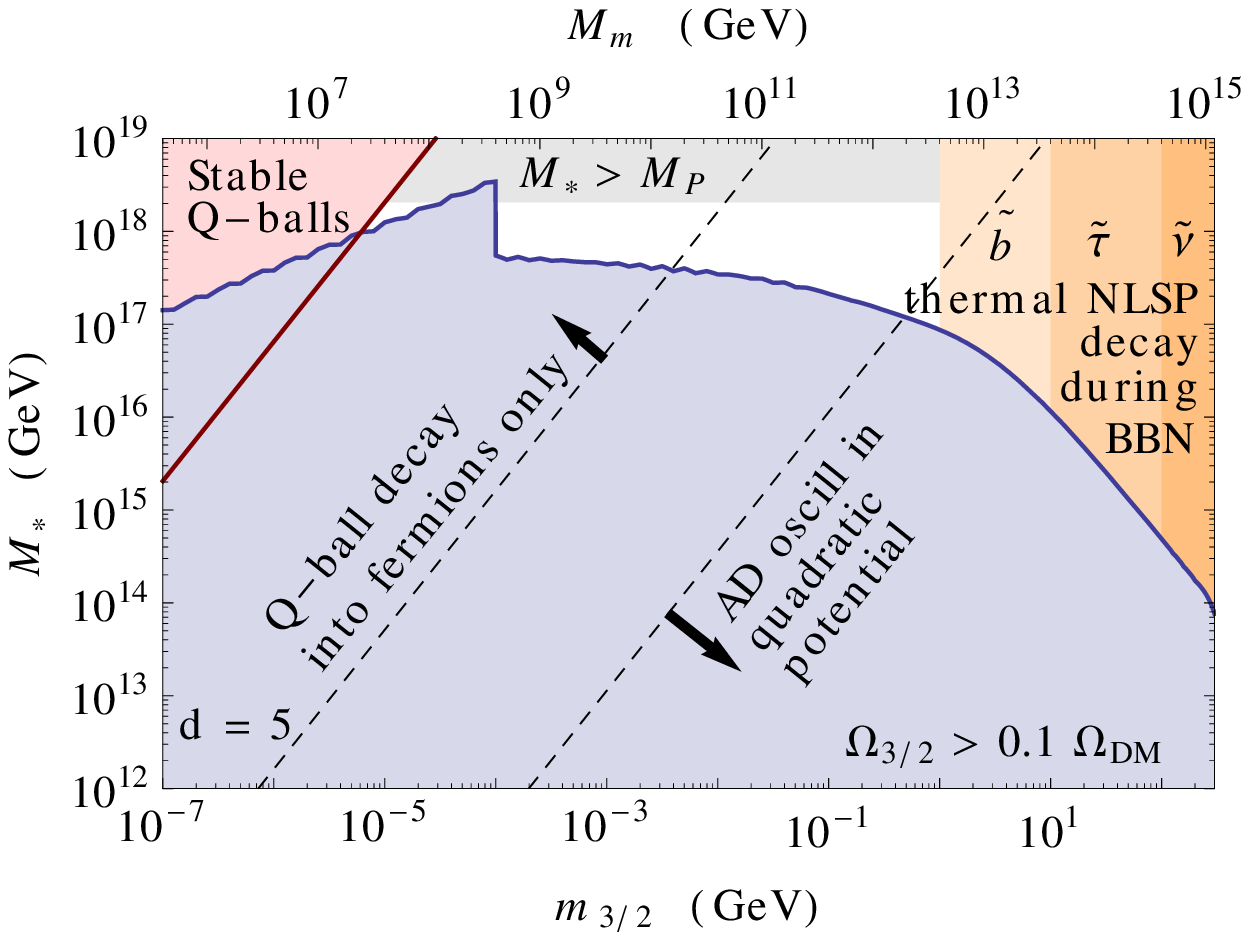}~~\includegraphics[width=7.2cm]{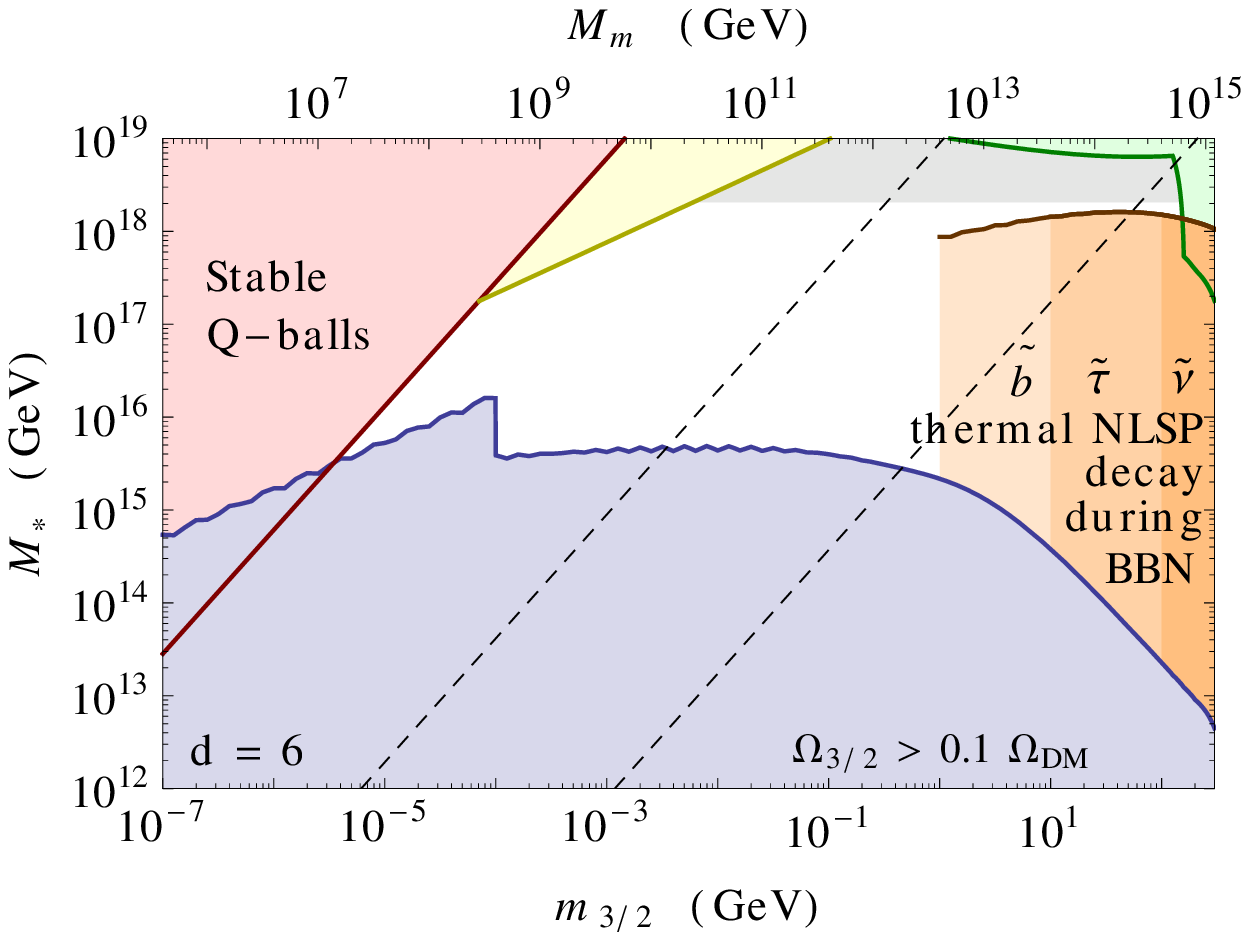}

\vspace{0.2cm}

\includegraphics[width=7.2cm]{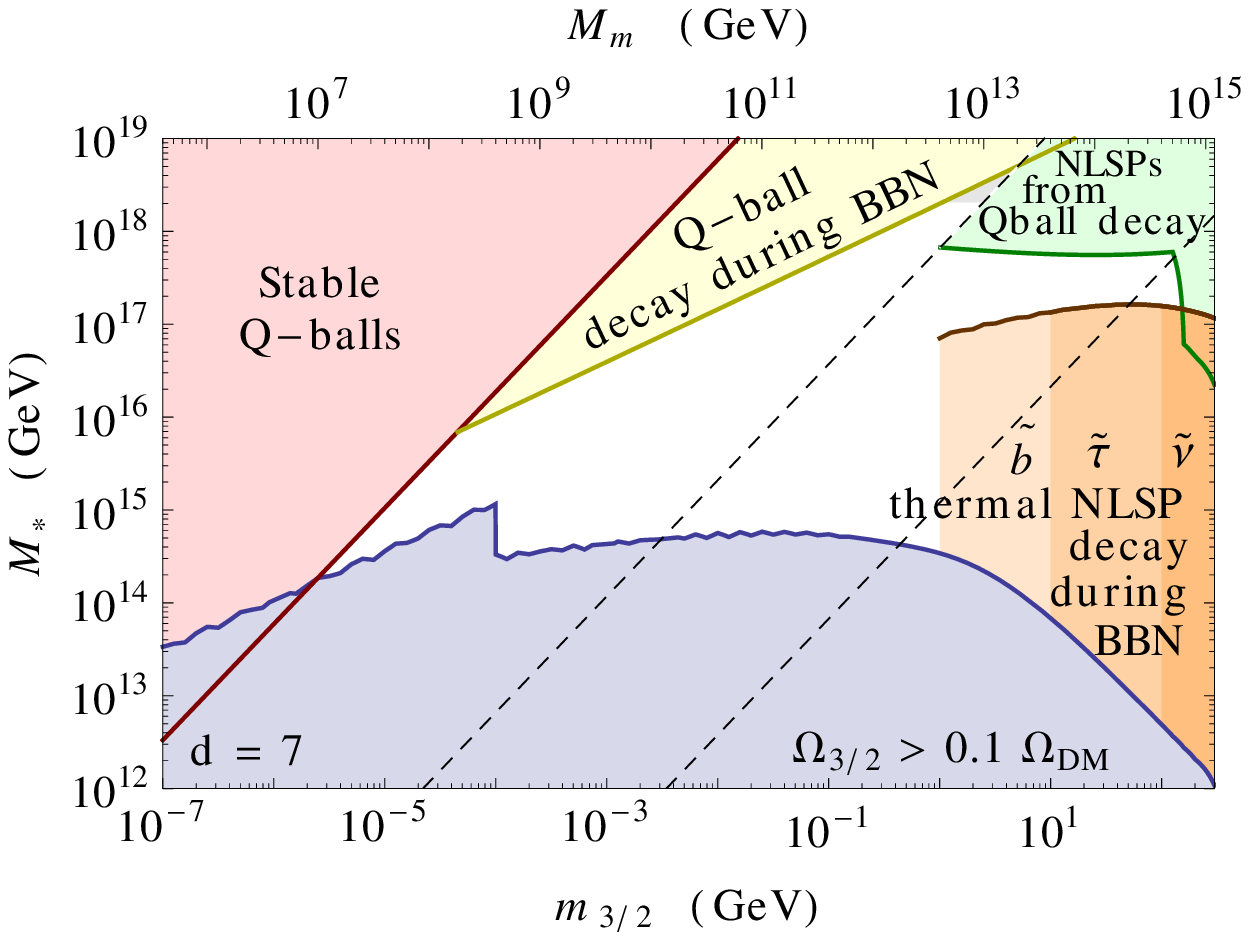}~~\includegraphics[width=7.2cm]{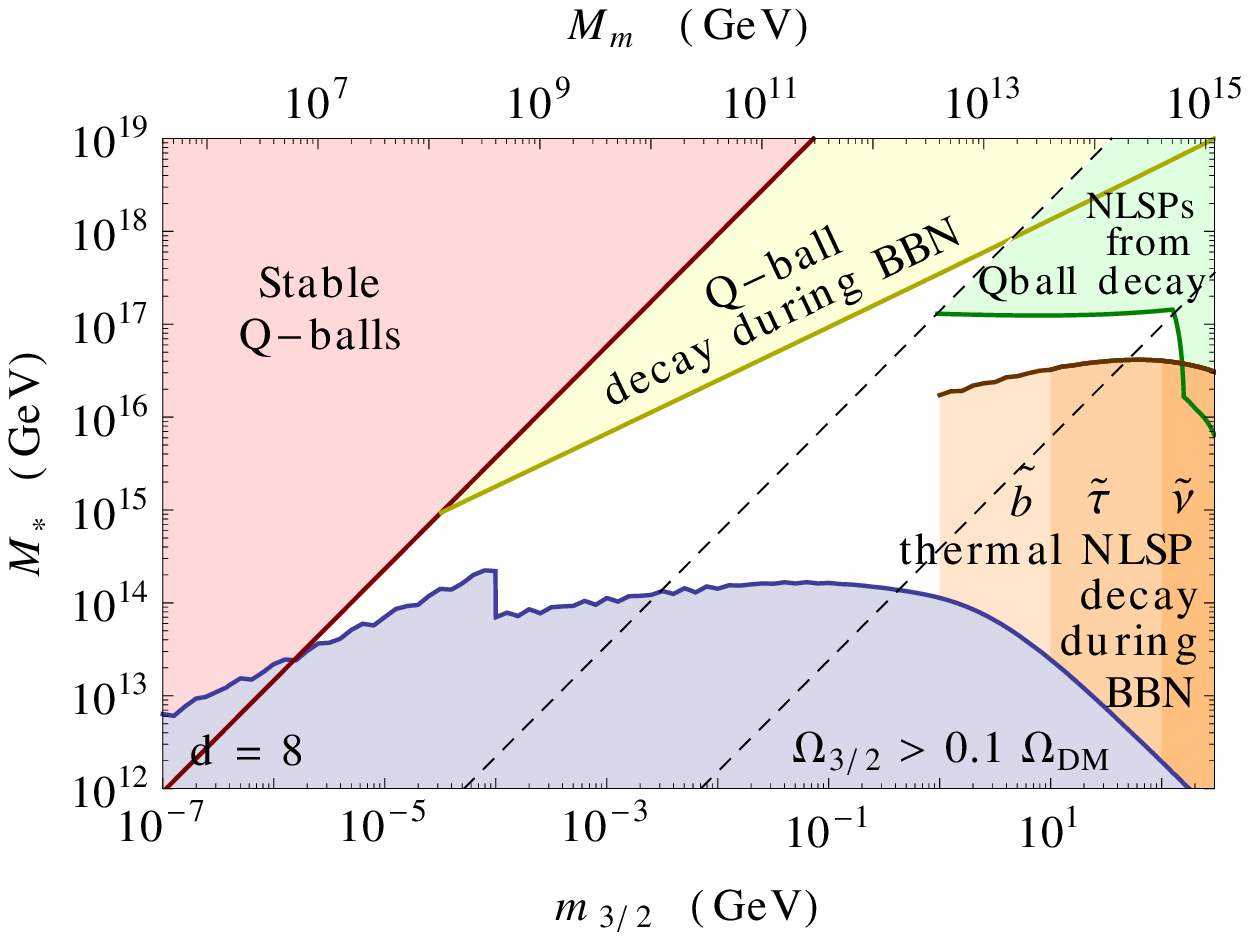}
\caption{
Various regimes on the $m_{3/2}-M_*$ or $M_m - M_*$ plane for asymmetry generation via the Affleck-Dine mechanism in GMSB models, for arbitrary messenger scale. Here, $m_{3/2}$ is the gravitino mass, $M_m$ is the messenger scale, and $M_*$ is the scale of the operator which lifts the flat direction.
The scenario of pangenesis is implemented as described in this paper, within the non-shaded areas for different values of the reheating temperature. \newline
The meaning of the shaded areas is as follows (cf. Eqs.~\eqref{eq:TR limit} and \eqref{eq:summ Qdec} -- \eqref{eq:summ b<1}): \newline
(i) In the blue area, gravitino DM is dominant or overabundant. \newline
(ii) In the orange areas, the NLSPs decay late and may spoil BBN. The upper bound on $m_{3/2}$ depends on the nature of the NLSP; it is at 1 GeV  for bino, 10 GeV for stau and 100 GeV for sneutrino NLSP. Above the orange line, the NLSPs do not thermalise while relativistic, their abundance is always suppressed, and their decay does not affect BBN. \newline
(iii) In the red area, the $Q$-balls formed after the fragmentation of the AD condensate are stable.  \newline
(iv)  In the yellow area, $Q$-balls decay at low temperatures $T_Q \lesssim 10 \text{ MeV}$, during or after BBN. \newline
(v)   In the green area, the gravitinos produced in the decay chain $Q\,\rm{ball} \to \rm{NLSP} \to \rm{gravitino}$ may become overabundant (we note that this is a stringent limit, and could be significantly relaxed).  \newline
(vi)  In the gray area, the scale of the operator that lifts the FD and produces the asymmetry exceeds the reduced Planck mass, $M_* >M_\rm{P}$. \newline
To guide the eye, we have drawn the lines $b=1$ and $b^{\frac{d-2}{d-1}}=70$, where $\smash{b \equiv \left(\ms M_*^{d-3} / M_m^{d-2} \right)^{1/(d-2)}}$. 
For $b\lesssim 1$ (dashed line on the right), the oscillations of the AD field  start in the quadratic part of the potential, and the $Q$-balls formed are gravity-type. For $b^{\frac{d-2}{d-1}}>70$ (dashed line on the left), $Q$-ball decay does not produce NLSPs (except at the late stages of evaporation). 
The four graphs correspond to different dimensionality of the operator which lifts the flat direction and is responsible for the U(1)$_X$ violation: $d=5$ upper left, $d=6$ upper right, $d=7$ bottom left, $d=8$ bottom right.
For these graphs, we have used $\ms = 500 \text{ GeV}, \ \eta_{_X} = 10^{-9}, \ q_{_X} = 1, \ \sin \delta = 1$ and $f_Q = 1$. }
\label{fig:m3/2--Mstar}
\end{figure}

The charge-to-entropy ratio, $\hX$, depends on four parameters: the reheating temperature $T_R$, the messenger mass $\Mm$, the scale $\Mnr$ at which the U(1)$_X$ violation occurs, and the dimension $d$ of the polynomial which lifts the FD. We take $\ms \sim 500 \GeV$ for definiteness. The requirement that $\hX \approx 10^{-9}$ reduces the free parameters to three. This means that for a fixed value of $d$, the required reheating temperature is a function of $\Mnr$ and $\Mm$, i.e. $T_R = \cal{\tilde{F}}_d(\Mnr,\Mm)$.
The reheating temperature is independently constrained, as summarised in \eq{eq:TR limit}, and the constraints depend on the gravitino mass. In GMSB, the gravitino mass is related to the messenger scale by
\beq
\Mm 
\approx \frac{\a}{4 \p} \pare{\frac{\mg \Mp}{\ms}} 
\approx 10^{-3} \pare{\frac{\mg \Mp}{\ms}} \ .
\label{eq:Mm}
\eeq
It is then convenient, by means of \eq{eq:Mm}, to express the reheating temperature that is required for the generation of sufficient asymmetry, in terms of $\Mnr$ and the gravitino mass, ${T_R = \cal{F}_d(\Mnr,\mg)}$.
This makes it possible to map the reheating-temperature constraints on the $\mg-\Mnr$ or the $\Mm-\Mnr$ plane, for various values of $d$. We do so in the graphs of  Fig.~\ref{fig:m3/2--Mstar}, where we also sketch various regimes associated with the production and decay of $Q$-balls, which we discuss in Sec.~\ref{sec:Q}.

\begin{figure}[t]
\centering
\includegraphics[width=7.6cm]{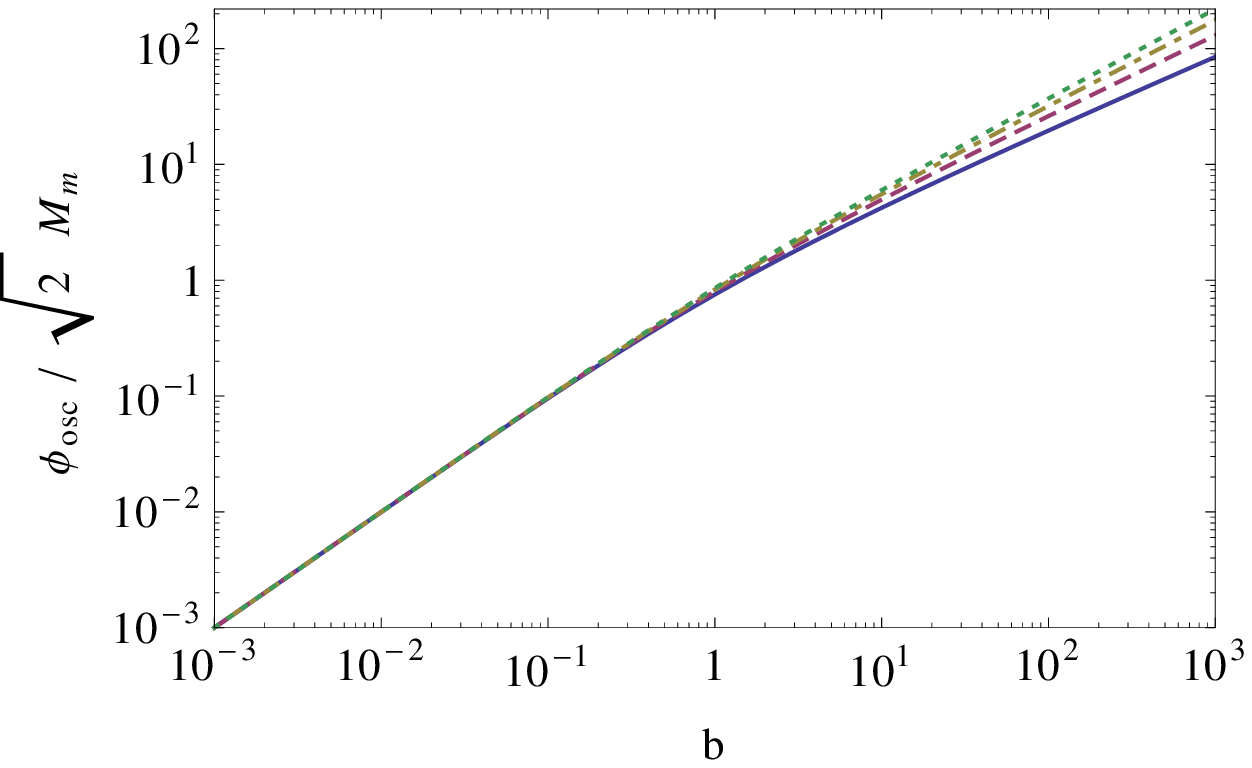}~\includegraphics[width=7.6cm]{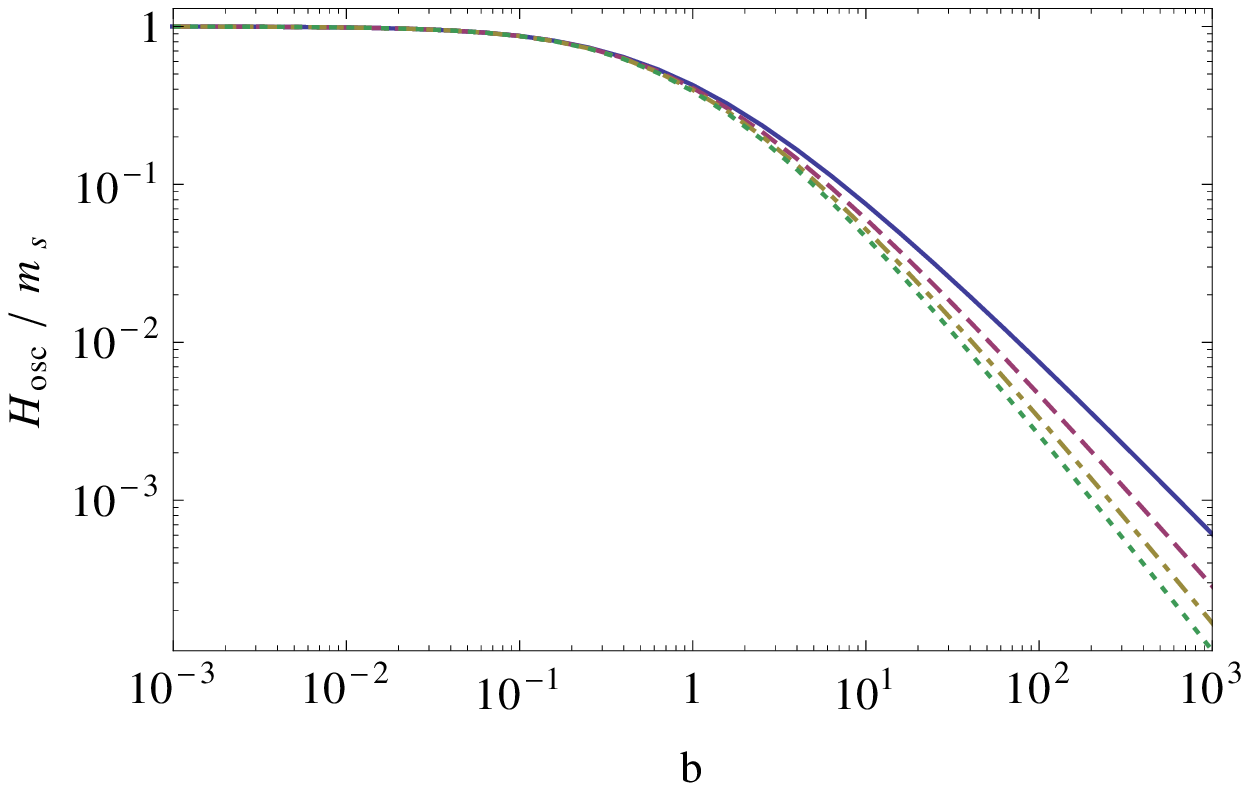}
\caption{The field amplitude over the messenger scale, $\phi_\rm{osc}/\sqrt{2}M_m$, and the Hubble parameter over the soft mass, $H_\rm{osc}/\ms$, at the onset of oscillations, plotted as a function of the dimensionless variable $b \equiv \left(\ms M_*^{d-3} / M_m^{d-2}\right)^{1/(d-2)}$ for the GMSB potential of Eq.~\eqref{eq:V GMSB}. The various curves correspond to different values of the dimension $d$ of the monomial which lifts the flat direction: $d=5$ (solid, blue), $d=6$ (dashed, purple), $d=7$ (dot-dashed, yellow), $d=8$ (dotted, green). The behaviour of the curves is described by Eqs.~\eqref{eq:sol b<1} and \eqref{eq:sol b>1}.}
\label{fig:fosc,hosc vs b}
\end{figure}

This requires to first determine the onset of oscillations of the AD field.
Equations~\eqref{eq:Hosc GMSB} and \eqref{eq:osc init} can be solved for the dimensionless variables $\fosc/\sqrt{2}\Mm$ and $\Hosc/\ms$ , in terms of the dimensionless parameter $b$, defined in \eq{eq:b}. We present the numerical solution in Fig.~\ref{fig:fosc,hosc vs b}, and we shall now describe the semi-analytical behaviour. For completeness, we also discuss why the thermal corrections of \eqs{eq:VT<} and \eqref{eq:VT>}, which were neglected in \eq{eq:Hosc GMSB}, are not relevant in the parameter region allowed by gravitino constraints. 

We discern the following cases:

\begin{list}{\labelitemi}{\leftmargin=1em}

\item For $b\lesssim 1$, the solution to Eqs.~\eqref{eq:Hosc GMSB} and \eqref{eq:osc init} is well approximated by
\beq
\frac{\fosc}{\sqrt{2} \Mm} \simeq b \ , \qquad  \frac{\Hosc}{\ms} \simeq 1 \ ,
\label{eq:sol b<1}
\eeq
and $A \approx a_o \ms + \mg$, where $a_o\sim 0.01$ is the loop-suppression factor of the $A$-term in GMSB.
Since $\fosc \lesssim \sqrt{2} \Mm$, the oscillations begin when the field is in the quadratic part of the potential.
The $A$-term is subdominant, and the ellipticity of the condensate is 
\beq \e_{_{\: b \, \lesssim 1}} \simeq \sin \delta  \, \frac{A}{\ms} \approx \sin \delta \left[ a_o + \pare{\frac{\mg}{\ms}} \right] < 1  \  . \label{eq:ellipt b<1} \eeq
As per \eq{eq:eta}, the charge-to-entropy ratio is
\beq
\hX \approx \frac{\qX \sin \delta}{2}   \pare{ \frac{A}{\ms} } b^2 \pare{\frac{\Mm^2 \, T_R}{\ms \, \Mp^2}} \ .
\label{eq:eta b<1}
\eeq

Thermal corrections:
We need to compare $\fosc$ with $\Tosc$. Let us first show that ${\fosc \gg \Tosc}$, thus the thermal corrections are of the type of \eq{eq:VT>} -- rather than \eq{eq:VT<} -- for reasonable values of the couplings $y$. 
In what follows, we will adopt the value ${\ms \approx 500 \GeV}$, to make the estimates more definite.

At $H= \Hosc \simeq \ms$, the temperature is $\Tosc \simeq (\ms T_R^2 \Mp)^{1/4}$, where $T_R$ is determined from \eq{eq:eta b<1} with the requirement that $\hX = 10^{-9}$. Setting  $\qX=\sin \delta = 1$, this gives
\beq 
\Tosc \approx 10^{21} \GeV^2 / \fosc  \  .
\label{eq:Tosc b<1}
\eeq 
The significance of the thermal corrections is maximal for minimal $\fosc$. We will now determine the minimum value of $\fosc$ allowed by the constraints on the reheating temperature. Combining \eqs{eq:TR limit} and \eqref{eq:eta b<1}, we find
\beq
\fosc \simeq \sqrt{2} \: b \Mm \gtrsim 
\left \{
\bal{4}
& 10^{12} \GeV \pare{\frac{1 \GeV}{\mg}}^{1/2}, &\quad &\rm{if} \quad \mg < \tilde{m} \\
& 4 \cdot 10^{14} \GeV , &\quad &\rm{if} \quad \mg > \tilde{m} \ ,
\eal
\right. 
\label{eq:con b<1}
\eeq
where $\tilde{m}$ is defined in \eq{eq:m3/2 con}. Note that because we are in the $b \lesssim 1$ regime, \eq{eq:con b<1} also implies that $\Mm \gtrsim 10^{12} \GeV$ or $\mg \gtrsim 0.5 \GeV$.

Comparing \eqs{eq:Tosc b<1} and \eqref{eq:con b<1}, we see that $\fosc \gg \Tosc$. We shall thus consider the thermal potential of \eq{eq:VT>}.
The thermal correction to $\Hosc^2$ arising from \eq{eq:VT>} is of the order ${[(y^2/4 \p) \: \Tosc^2/\fosc]^2}$. Let us now compare this to the value of $\Hosc \simeq \ms$ given in \eq{eq:sol b<1}. Using the above, we get
\beq
\frac{y^2}{4 \p} \frac{\Tosc^2/\fosc}{\ms} < \left \{
\bal{4}
& 10^2 y^2  \pare{\frac{\mg}{1 \GeV}}^{3/2}, & \quad &\rm{if} \quad \mg < \tilde{m}   \\ 
& 10^{-6} y^2 ,                              & \quad &\rm{if} \quad \mg > \tilde{m}   \, .
\eal
\right.
\eeq
Thus we find that, if $\mg < \tilde{m}$, the thermal contribution to $\Hosc$  is subdominant at least for $\smash{y \lesssim 0.1 (1\GeV / \mg)^{3/4}}$. If $\mg > \tilde{m}$, on the other hand, the thermal correction to $\Hosc$ is obviously irrelevant.

\medskip

\item For $1 < b < \pare{\ms / \mg}^{ \frac{d-1}{d-2} }$, the solution to Eqs.~\eqref{eq:Hosc GMSB} and \eqref{eq:osc init} is well approximated by
\beq
\frac{\fosc}{\sqrt{2}\Mm} \simeq  b^{ \frac{d-2}{d-1} } \ , \qquad  
\frac{\Hosc}{\ms} \simeq  b^{- \frac{d-2}{d-1} }  \ .
\label{eq:sol b>1}
\eeq
Since $\fosc \gtrsim \sqrt{2} \Mm$, the oscillations begin when the field is in the flat part of the potential. 
Using the above solution, the parameter $A$ of \eq{eq:A GMSB} becomes
\beq
A \simeq \mg + a_o \ms b^{- \frac{d-2}{d-1} }\ .
\eeq
The ellipticity of the condensate then is
\beq 
\e_{_{\: b \, > 1}} = \sin \delta \frac{A}{\Hosc}  \approx \sin \delta \left[ a_o  + \pare{\frac{\mg}{\ms}} b^\frac{d-2}{d-1}  \right]  \  ,
\label{eq:ellipt b>1}
\eeq 
and remains less than unity since $b^\frac{d-2}{d-1} < \ms / \mg$ in this regime. Using \eq{eq:eta}, the $X$-charge-to-entropy ratio is
\beq
\hX \approx \frac{\qX \sin \delta}{2} \pare{ \frac{A}{\ms} } b^{\frac{4(d-2)}{d-1}} \pare{\frac{\Mm^2 \, T_R}{\ms \, \Mp^2}} \ .
\label{eq:eta b>1}
\eeq

Thermal corrections: We shall again compare $\fosc$ with $\Tosc$. At $H = \Hosc \simeq \ms \pare{\Mm/\fosc}$, the temperature $\Tosc \approx (\Hosc T_R^2 \Mp)^{1/4}$ is
\beq
\Tosc \approx 10^{21} \GeV^2 \: \pare{ \Mm^3 / \fosc^7 }^{1/4} \approx 4 \cdot 10^{30} \GeV^2 \: \pare{ \mg^3 / \fosc^7 }^{1/4} \, ,
\label{eq:Tosc b>1}
\eeq
where we solved \eq{eq:eta b>1} for $T_R$, requiring as before that $\hX = 10^{-9}$. Using the constraints of \eq{eq:TR limit} on $T_R$, we find that $\fosc$ must satisfy
\beq
\fosc \simeq \sqrt{2} \: b^\frac{d-2}{d-1} \: \Mm \gtrsim \left \{
\bal{4}
& 10^{12} \GeV, &\quad &\rm{if} \quad \mg < \tilde{m} \\
& 10^{14} \GeV \pare{\frac{\mg}{1 \GeV}}^{1/3} , &\quad &\rm{if} \quad \mg > \tilde{m} \ .
\eal
\right. 
\label{eq:con b>1}
\eeq
Equations \eqref{eq:Tosc b>1} and \eqref{eq:con b>1} can be combined to show that $\fosc/\Tosc \gg 1$ for the range of gravitino masses of interest. Thus, we shall again consider the thermal correction to $\Hosc$ arising from \eq{eq:VT>}, and compare it to the value of $\Hosc$ given in \eq{eq:sol b>1}. Using the above, we get
\beq
\frac{y^2}{4 \p} \frac{\Tosc^2 / \fosc}{\Hosc} 
\simeq y^2 \pare{\frac{6 \cdot 10^{12} \GeV}{\fosc}}^{7/2} \pare{\frac{\mg}{1 \GeV}}^{1/2}  \  .
\eeq
Applying \eq{eq:con b>1}, we find
\beq
\frac{y^2}{4 \p} \frac{\Tosc^2 / \fosc}{\Hosc} 
\lesssim \left\{
\bal{4}
& 600 \, y^2  \pare{\frac{\mg}{1 \GeV}}^{1/2}  , & \quad &\rm{if} \quad \mg < \tilde{m}   \\ 
& 6 \cdot 10^{-5} \, y^2  \pare{\frac{1 \GeV}{\mg}}^{2/3} ,                              & \quad &\rm{if} \quad \mg > \tilde{m}   \, .
\eal
\right.
\eeq
If $\mg < \tilde{m}$, the thermal contribution is subdominant at least for $y \lesssim 0.01$, and may only enhance $\Hosc$ by a factor of a few for $0.1 \lesssim y \lesssim 1$.
For $\mg > \tilde{m}$, on the other hand, the thermal correction to $\Hosc$ is completely negligible for the allowed values of $T_R$.

\item For $b \gtrsim \pare{\ms / \mg}^{ \frac{d-1}{d-2}}  \gg 1$, the oscillations begin in a region of the potential where the contribution from PMSB is relevant: The corresponding term in the potential is ${V_\rm{PMSB} \sim \mg^2 |\F|^2}$ (cf.~\eq{eq:V PMSB}).
Comparing with the first term in \eq{eq:V GMSB}, we see that PMSB effects are important at the onset of oscillations if $\mg^2 \fosc^2 \gtrsim \ms^2 \Mm^2$. From \eq{eq:sol b>1} (which is valid up to the overlapping region between the two regimes), we find that this happens for $b \gtrsim \pare{\ms / \mg}^{ \frac{d-1}{d-2}}$ or, equivalently,
${\pare{\mg \Mnr^{d-3}}^{1/(d-2)} > 10^{-3}\Mp}$. Requiring that $\Mnr< \Mp$ and $\mg < \ms$, this regime is only possible for $d \geq 8$. We will restrict ourselves to $d \leq 8$ when presenting the allowed regions of parameter space and have checked that this regime is excluded by constraints from $Q$-ball decay for $d=8$ (see Fig.~\ref{fig:m3/2--Mstar}). We shall therefore not consider this case further.

\end{list}

Using the above analysis of the dynamics which generates the asymmetry, we numerically translate the bounds of \eq{eq:TR limit} into exclusion regions on the $\mg - \Mnr$ plane, for $d=5-8$ and present the results in Fig.~\ref{fig:m3/2--Mstar}. For each parameter pair in the allowed regions, the $X$-charge-to-entropy ratio $\hX \simeq 10^{-9}$ can be obtained for a value of the reheating temperature which satisfies the bounds in \eq{eq:TR limit}. Additional constraints for successful pangenesis arise from considering the late-time evolution of the AD condensate, which we now discuss.

\subsection{Formation and decay of $Q$-balls}
\label{sec:Q}

The Affleck-Dine condensate is unstable with respect to spatial perturbations~\cite{Kusenko:1997si,Kusenko:1997ad,Kusenko:1997zq,Enqvist:1997si,Enqvist:1998en,Enqvist:2000gq,Kasuya:1999wu,Kasuya:2000wx,Kasuya:2001hg,Hiramatsu:2010dx}. If the potential of the AD field grows more slowly than $\f^2$, the AD condensate can fragment into bound states of the scalar fields participating in the flat direction, known as $Q$-balls~\cite{Coleman:1985ki}. $Q$-balls are non-topological solitonic configurations of fields which transform under a global U(1) symmetry. They carry a non-zero charge under this symmetry, and they are stable against decay into quanta of the same field. If they form, they may affect the cascade of the charge generated in the condensate down to the lightest visible and dark-sector baryons.

$Q$-balls can decay into gauge-invariant combinations of particles which carry the same charge. They may decay into scalar degrees of freedom that are linearly independent from the $Q$-ball field, and/or fermions. A given $Q$-ball decay channel is kinematically allowed only if the mass-to-charge ratio of the $Q$-ball is larger than that of the decay products. 
The decay into fermions proceeds only from the surface of the $Q$-ball, while the decay into scalars can occur from the interior. The latter is thus typically enhanced with respect to the former. We parametrise this enhancement of the decay rate into scalars by a factor $f_Q$ which can be as large as $f_Q \lesssim 10^3$, depending on the flat direction~\cite{Enqvist:1998en}. For decays to fermions, we correspondingly set $f_Q=1$. The decay rate of $Q$-balls has an upper limit of~\cite{Cohen:1986ct}
\beq
\G_Q \approx f_Q \frac{\w_Q^3}{48 \p} \frac{ R_Q^2}{Q}  \ ,
\label{eq:GammaQ}
\eeq
where $\w_Q$ is the rotational speed of the $Q$-ball field configuration, $\F_Q = \sqpare{\f_Q(r)/\sqrt{2}} \exp(i \w_Q t)$, in the internal phase space, $R_Q$ is the radius and $Q$ is the charge of the $Q$-ball. Note that the charge is measured in units of $\f$ quanta and is related to the $X$ charge by
\beq
Q_X = \qX Q
\label{eq:QX}
\eeq
where $\qX$ is the $X$ charge assignment of the FD, defined in \eq{eq:FD qX}. The upper limit on the decay rate is reached if $y \f_Q / \w_Q \gg 1$, where $y$ is the coupling of the $Q$-ball field to the fields it decays into. The decay rate is typically saturated for $Q$-balls formed via fragmentation of an AD condensate and which couple to visible and dark sector fields (see e.g.~\cite{Kasuya:2011ix}).
The quantities $R_Q, \ \w_Q$ and $\f_Q$ depend on the parameters of the low-energy AD potential and the charge $Q$. However, the charge $Q$ itself is determined from the fragmentation of the AD condensate, and thus depends on the dynamics of the AD mechanism at higher energies.

The fragmentation of an AD condensate and the formation of $Q$-balls has been studied semi-analytically and numerically. In what follows, we use these studies to delineate various regimes in the parameter space of the AD potential, with respect to the formation, stability or lifetime, and decay modes of the $Q$-balls. We depict these regimes in Fig.~\ref{fig:m3/2--Mstar}.

For the purposes of the pangenesis scenario described in this paper, we shall focus on the regions where if $Q$-balls form, they are unstable and decay sufficiently early, releasing all of their charge to the thermal plasma without spoiling any of the predictions of the standard cosmology. To be conservative, we shall require that the $Q$-balls formed after Affleck-Dine pangenesis decay before BBN, the latest at temperature
\beq
T_Q \gtrsim 10 \MeV  \  .
\label{eq:TQ limit}
\eeq
We outline this region in the graphs of Fig.~\ref{fig:m3/2--Mstar}, using the properties of $Q$-balls described in the following subsections. Additional stronger requirements may arise, depending on the type of $Q$-ball, which we shall now discuss.

\subsubsection{$Q$-balls in gravity-mediated SUSY breaking}
\label{sec:Q-PMSB}

After the onset of oscillations, the dominant part of the FD potential for the case of PMSB in \eq{eq:V PMSB} is
\beq
V_\rm{AD} \supset  \ms^2 |\F|^2 \sqpare{1 + K \ln \pare{ \frac{|\F|^2}{\Mp^2} } } \  ,  \nn
\label{eq:PMSB--V low}
\eeq
with $\ms \sim \mg \sim 500 \GeV$.
The formation of $Q$-balls depends critically on the sign of the parameter $K$:
If $K>0$, the potential grows faster than quadratic, and no $Q$-balls are formed.
If $K<0$, the potential grows more slowly than quadratic, and $Q$-ball solutions exist.

The correction proportional to $K$ arises at 1-loop, and typically $|K| \sim 0.01 - 1$.
Gaugino loops contribute $\delta K<0$, while Yukawa couplings give $\delta K>0$~\cite{Enqvist:2000gq}. 
In the MSSM, $K>0$ arises typically for purely leptonic FDs (since they do not couple to the gluino), and for baryonic FDs with large third-generation contribution (as long as the gluino is lighter than the stops). For baryonic FDs of first and second generation fields, $K<0$.

If $K<0$, most of the charge density carried initially by the condensate gets trapped in $Q$-balls. The average charge of these $Q$-balls can be estimated from numerical simulations as~\cite{Hiramatsu:2010dx}
\beq
Q 
\ \approx \ 2 \cdot 10^{-2} \: \pare{ \frac{\fosc}{\ms} }^2 
\ \approx \ 4 \cdot 10^{-2} \: \pare{ \frac{\Mnr}{\ms} }^{\frac{2(d-3)}{d-2}} \ ,
\label{eq:PMSB--Q} 
\eeq
where we used \eq{eq:fmin} and the fact that $\Hosc \sim \ms$ in PMSB, and assumed ellipticity $\e \approx 1$. For the potential in PMSB, the mass, radius and rotational speed of a $Q$-ball with charge $Q$ are~\cite{Enqvist:1997si,Enqvist:1998en}
\begin{align}
 M_Q  &= \ms Q  \label{eq:PMSB--MQ} \ , \\
 R_Q  &= \frac{\sqrt{2}}{|K|^{1/2} \ms}  \label{eq:PMSB--RQ} \ , \\
 \w_Q &= \ms  \label{eq:PMSB--wQ}  \ .
\end{align}
Using the above and \eq{eq:GammaQ}, the temperature at the time of $Q$-ball decay is
\beq
T_Q  = \frac{5^{1/4}}{2^{5/4} \p}  \pare{\frac{f_Q \ms \Mp}{|K| Q}}^{1/2}  \label{eq:PMSB--TQ} \ .
\eeq
For the charge of \eq{eq:PMSB--Q}, we then find
\beq
T_Q 
\ \approx \
\pare{\frac{f_Q \: \ms \Mp}{|K|}}^{1/2}
\pare{\frac{\ms}{\Mnr}}^\frac{d-3}{d-2} 
\ \approx \ 
\pare{ \frac{f_Q}{2 |K|} \: \frac{\qX \sin \delta}{\hX} \: \frac{T_R}{\Mp}}^{1/2}  \ms \  ,
\label{eq:PMSB--TQ num}
\eeq
where in the second equality we used \eq{eq:eta} to express $\Mnr$ in terms of the reheating temperature $T_R$ (and the $X$-charge-to-entropy ratio $\hX$).

The mass-to-charge ratio of $Q$-balls in PMSB, $M_Q/Q \sim \ms$, allows them to decay into fermions and generically also into scalars (unless the FD field consists only of the LSP).  
If $Q$-balls decay after the LSP freezes out, they will introduce a non-thermal LSP density, which has to be subdominant for successful pangenesis. If this density is sufficient to bring the LSPs in equilibrium, annihilations will reduce it to
\beq
n_\rm{LSP}^Q(T_Q)  \approx \sqpare{\frac{H}{\ang{\s v}}}_{T=T_Q}    \ .
\label{eq:LSP equil}
\eeq
The abundance determined by \eq{eq:LSP equil} is the upper limit on the LSP relic density from $Q$-ball decays. We emphasise that the actual abundance may be smaller.
It is useful to compare $n_\rm{LSP}^Q(T_Q)$ with the LSP relic density from thermal freeze-out (which has to be a subdominant component of DM in models of pangenesis, as we discuss in Sec.~\ref{sec:models}). At temperature $T=T_Q$, the latter is
\beq
n_\rm{LSP}^f(T_Q) \simeq n_\rm{LSP}^f(T_f) \: \frac{T_Q^3}{T_f^3} \simeq \frac{T_Q^3}{T_f^3} \sqpare{\frac{H}{\ang{\s v}}}_{T=T_f}  ,
\label{eq:thermal relic}
\eeq
where $T_f$ is the freeze-out temperature of the LSP. Combining \eqs{eq:LSP equil} and \eqref{eq:thermal relic}, we get
\beq
\frac{\W_\rm{LSP}^Q}{\W_\rm{LSP}^f}
= \frac{n_\rm{LSP}^Q (T_Q)}{n_\rm{LSP}^f  (T_Q)}
\approx  \frac{T_f}{T_Q} \: \frac{\ang{\s v}_f}{\ang{\s v}_Q} 
\approx \frac{T_f}{T_Q} \ \rm{or} \ \frac{T_f^2}{T_Q^2} \ ,
\label{eq:W LSP Q}
\eeq
for $s$- or $p$-wave annihilation of the LSPs. If $\W_\rm{LSP}^f \ll \W_\rm{DM}$, it is possible to accommodate $Q$-ball decay after the LSP freeze-out while maintaining subdominant LSP abundance. For simplicity and in order to avoid any extra assumptions, however, we shall require $T_Q \gtrsim T_f$. Typically, $T_f \approx m_\rm{LSP} / 20$, and we thus impose
\beq T_Q \gtrsim T_f \sim 10 \GeV \ . \label{eq:TQ 10GeV} \eeq

Given \eq{eq:PMSB--TQ num}, the constraint of \eq{eq:TQ 10GeV} poses an upper limit on $\Mnr$ or, equivalently, a lower limit on $T_R$:
\beq
T_R \gtrsim 10^4 \GeV \ \frac{1 }{f_Q} \pare{\frac{|K|}{0.01}} \pare{\frac{500 \GeV}{\ms}}^2  \pare{\frac{T_f}{10 \GeV}}^2  \ .
\label{eq:TR limit PMSB}
\eeq
We denote this limit in Fig.~\ref{fig:PMSB TR vs Mnr} (for the conservative choice of $f_Q\sim 1$). We note, however, that this is a worst-case-scenario limit if $Q$-balls indeed form ($K<0$), which does not apply if $Q$-balls do not form ($K>0$).

\subsubsection{$Q$-balls in gauge-mediated SUSY breaking}
\label{sec:GMSB Qballs}

After the onset of oscillations, the dominant part of the FD potential for the case of GMSB in \eq{eq:V GMSB} is
\beq
V_\rm{AD} =  \ms^2 \Mm^2 \ln^2 \pare{1+ \frac{|\F|}{\Mm}} \sqpare{1 + K \ln \pare{ \frac{|\F|^2}{\Mm^2} } }  \ . \nn
\label{eq:GMSB--V low}
\eeq
Following the analysis of Sec.~\ref{sec:GMSB param}, we discern two cases: $b \lesssim 1$ and $1 < b < \pare{\ms / \mg}^{ \frac{d-1}{d-2} }$, where $b \equiv \pare{\ms \Mnr^{d-3}/\Mm^{d-2}}^{1/(d-2)}$, as defined in \eq{eq:b}.

\medskip

If $b \lesssim 1$, the oscillations of the AD field start in the quadratic part of the potential, at $\fosc/\sqrt{2} \simeq b \Mm$, and the situation resembles the PMSB case. $Q$-balls form only if $K<0$. If they form, most of the charge density of the condensate is carried by $Q$-balls of charge~\cite{Hiramatsu:2010dx}
\beq
Q_{\: b \, \lesssim 1} 
\ \approx \ 10^{-3} \pare{\frac{\fosc}{\ms}}^2 \\
\ \approx \ 10^{17} \: b^2 \pare{\frac{\mg}{1 \GeV}}^2 \pare{\frac{500 \GeV}{\ms}}^4 \ ,
\label{eq:GMSB--Q b<1}
\eeq
where the prefactor is suppressed with respect to \eq{eq:PMSB--Q} due to the lower ellipticity of the condensate, $\e < 1$, in the GMSB case. The other properties of the $Q$-balls are as described in the previous subsection, \eqs{eq:PMSB--MQ} to \eqref{eq:PMSB--TQ}. Substituting \eq{eq:GMSB--Q b<1} into \eqref{eq:PMSB--TQ}, we find the temperature at the time of $Q$-ball decay
\beq
T_{Q, \: b \lesssim 1} \
\approx \ 250 \GeV \: \frac{1}{b}  \pare{\frac{1 \GeV}{\mg}} \: f_Q^\frac{1}{2} \pare{\frac{0.01}{|K|}}^\frac{1}{2}  \pare{\frac{\ms}{500 \GeV}}^\frac{5}{2} \, .
\label{eq:GMSB--T b<1}
\eeq

Because of their large mass-to-charge ratio, $M_Q / Q \sim \ms$, the decay of these $Q$-balls produces NLSPs, which subsequently decay into gravitinos and $R$-parity even fermions. We must ensure that the gravitinos remain a subdominant component of DM. A calculation analogous to that of \eqs{eq:LSP equil} to \eqref{eq:W LSP Q} gives the maximum gravitino relic density as
\beq
\frac{\W_{3/2}^Q}{\W_\rm{NLSP}^f} 
\approx \frac{\mg}{m_\rm{NLSP}} \: \frac{T_f}{T_Q} \: \frac{\ang{\s v}_f}{\ang{\s v}_Q} 
\approx \frac{\mg}{\ms} \times \sqpare{\frac{T_f}{T_Q} \ \rm{or} \ \frac{T_f^2}{T_Q^2}} \ ,
\label{eq:W 3/2 Q}
\eeq
where $\W_\rm{NLSP}^f$ is the would-be NLSP thermal-relic density if NLSPs did not decay, and the two cases in brackets are for $s$- and $p$-wave annihilation, respectively.
Typically $T_f \approx \ms/20$. Assuming  $\W_\rm{NLSP}^f \sim \W_\rm{DM}$ and requiring $\W_{3/2}^Q < \W_\rm{DM}$, we get a rough constraint on $T_Q$. For the case of $s$-wave annihilation, this gives
\beq
T_Q \gtrsim \mg / 20  \  .
\label{eq:mg/20}
\eeq
Given \eq{eq:GMSB--T b<1}, this constraint can be written in terms of $b$ and $\mg$ as
\beq
b \pare{\frac{\mg}{1 \GeV}}^2  \lesssim 5 \cdot 10^3  \: f_Q^\frac{1}{2} \pare{\frac{0.01}{|K|}}^\frac{1}{2}  \pare{\frac{\ms}{500 \GeV}}^\frac{5}{2}  \ . 
\label{eq:Qdec con b<1}
\eeq
Since $b \lesssim 1$, this constitutes a rather mild upper limit on $\mg$, even for $f_Q =1$.

\medskip

For the case $1 < b < \pare{\ms / \mg}^{ \frac{d-1}{d-2} }$, the oscillations of the AD field start in the nearly-flat part of the potential, $\smash{\fosc / \sqrt{2}\Mm \simeq b^\frac{d-2}{d-1} > 1}$. Numerical simulations show that most of the charge density is trapped inside $Q$-balls of average charge~\cite{Kasuya:2001hg}
\beq
Q_{\: b \, > 1} \approx \b  \pare{\frac{\fosc^2}{\ms \Mm}}^2 \ ,
\label{eq:GMSB--Q} 
\eeq
where $\b$ depends on the ellipticity of the condensate: $\b \approx 6 \cdot 10^{-4}$ for $\e \simeq 1$ and  $\b \approx 6 \cdot 10^{-5}$ for $\e \lesssim 0.1$. Thus
\beq
Q_{\: b \, > 1} \approx 10^{16} \: b^\frac{4(d-2)}{d-1} \pare{\frac{\mg}{1 \GeV}}^2 \pare{\frac{500 \GeV}{\ms}}^4 \ .
\label{eq:GMSB--Q num} 
\eeq
Note that the two formulas for the Q-ball charges in the GMSB case, \eqs{eq:GMSB--Q b<1} and \eqref{eq:GMSB--Q num}, agree reasonably well in the overlapping region of validity $b \approx 1$, given that they are based on two different numerical simulations (Refs.~\cite{Hiramatsu:2010dx} and \cite{Kasuya:2001hg}, respectively).
The mass, radius, rotational speed of the field and field amplitude at the center of $Q$-balls which form in the nearly-flat part of the potential are~\cite{Kusenko:1997ad,Kusenko:1997zq,Dvali:1997qv,Kusenko:1997si,Kusenko:2005du}
\begin{align}
 M_Q  &= \frac{4 \sqrt{2} \p}{3} \pare{\ms \Mm}^{1/2} Q^{3/4}  \label{eq:GMSB--MQ} \ , \\
 R_Q  &= \frac{1}{\sqrt{2}} \frac{Q^{1/4}}{\pare{\ms \Mm}^{1/2}}  \label{eq:GMSB--RQ} \ , \\
\w_Q  &= \sqrt{2} \p \frac{\pare{\ms \Mm}^{1/2}}{Q^{1/4}}  \label{eq:GMSB--wQ} \ , \\
\f_Q  &= \pare{\ms \Mm}^{1/2} Q^{1/4} \label{eq:GMSB--fQ} \ .
\end{align}
The mass-to-charge ratio of this type of $Q$-ball thus decreases with charge:
\beq
\bal{2}
\frac{M_Q}{Q} \
&\approx \ \frac{4 \sqrt{2} \p}{3} \: \frac{\pare{\ms \Mm}^{1/2}}{Q^{1/4}}  \\
&\approx \ 2 \cdot 10^4 \GeV  \ b^{-\frac{d-2}{d-1}}  \pare{\frac{\ms}{500 \GeV}} \ .
\eal
\label{eq:GMSB--M/Q}
\eeq
For sufficiently large $Q$, the mass-to-charge ratio is smaller than the lightest gauge-invariant combination of quanta carrying unit charge. In this case, $Q$-balls are energetically disallowed to decay, and in many scenarios found in the literature they constitute (part of) the DM of the universe (see e.g.~\cite{Kusenko:1997si,Kasuya:1999wu,Demir:2000gj,Kusenko:2001vu,Kasuya:2002tu,Kusenko:2004yw,Barnard:2011jw}). $Q$-ball dark matter carrying (visible) baryonic charge is viable as long as the $Q$-balls were formed along FDs lifted by visible-baryon-number violating operators~\cite{Kusenko:2005du}.

For the pangenesis scenario described here (see Sec.~\ref{sec:models}), we require that the $Q$-balls formed can decay, so that all of the asymmetry generated in the AD condensate is released into the plasma, and cascades down to the lightest visible and dark sector particles (we will explore the possibility of antibaryonic $Q$-ball dark matter in the framework of pangenesis elsewhere). This sets an upper bound on the charge of the $Q$-balls formed. A more stringent bound arises from requiring $Q$-balls to decay sufficiently early. In this regime, the temperature at the time of $Q$-ball decay is
\beq
\bal{2}
T_{Q, \: b > 1} \
&\approx \ \sqpare{ \frac{\sqrt{5} \p \: f_Q}{8} \frac{(\ms \Mm)^{1/2} \Mp}{Q^{5/4}} }^{1/2} \\
&\approx \ 10^3 \GeV \  b^{-\frac{5(d-2)}{2(d-1)}} \pare{\frac{1 \GeV}{\mg}} f_Q^\frac{1}{2} \pare{\frac{\ms}{500 \GeV}}^\frac{5}{2}  \  .
\eal
\label{eq:GMSB--T b>1}
\eeq
The minimum temperature at which the $Q$-balls decay without changing the predictions of standard cosmology depends on their decay products. 
These are determined by the mass-to-charge ratio $M_Q/Q$. There are the following possibilities:
\begin{list}{\labelitemi}{\leftmargin=1em}
\item 
Large $Q$-balls with mass-to-charge ratio
\beq 
\frac{1}{\qX} \frac{M_Q}{Q} \ < \ \frac{ \sum_f m_f }{ \sum_f  q_{_{X,f}} } \sim 3 \GeV  
\label{eq:M/Q small}
\eeq
are stable. Here $\sum_f m_f$ is the sum of masses of the lightest gauge-singlet combination of fields which carry a net $X$ charge  $\sum_f  q_{_{X,f}}$. 
In pangenesis, this combination involves both visible and dark sector fields with equal $X$ charges, $q_{_{X,\rm{vis}}} =  q_{_{X,\rm{dark}}}$, such that a $(B-L)$-singlet combination is formed.
The lightest dark-sector combination of fields carrying ${X \ne 0}$ charge, and zero net charge under any dark-sector forces (except $B-L$), is the DM state itself.
As we shall see in Sec.~\ref{sec:models}, in order to get the right relic DM \emph{mass} density, the mass of the DM state (which may be composite) has to be ${m_\dm \simeq 5 \GeV \times  q_{_{X,\rm{DM}}}}$ (cf. \eq{eq:mDM gen}, for asymmetry release in the thermal bath below the electroweak phase transition). 
The visible-sector mass contribution to the right-hand side of \eq{eq:M/Q small} depends on the FD along which pangenesis occurs: it is negligible for a flat direction carrying $\Lv$ but no $\Bv$, and is $m_p \times q_{_{X,\rm{vis}}} \simeq 1 \GeV \times  q_{_{X,\rm{vis}}}$ for a flat direction carrying $\Bv$ but no $\Lv$. 
The above considerations yield a $(2.5 - 3) \GeV$ bound, as denoted in \eq{eq:M/Q small}.

The inequality \eqref{eq:M/Q small} is satisfied for
\beq
b^\frac{d-2}{d-1} > 10^4 \pare{ \frac{\ms}{500 \GeV} } \frac{1}{\qX}  \  ,
\label{eq:b bigQ}
\eeq
where we used \eq{eq:GMSB--M/Q}. As per \eq{eq:GMSB--Q num}, $Q$-balls with charge larger than
\beq
Q > Q_\rm{stable} \approx  10^{32} \pare{\frac{\mg}{1 \GeV}}^2 \frac{1}{\qX^4}
\label{eq:Qstable}
\eeq
are thus completely stable. We mark this regime in the plots of Fig.~\ref{fig:m3/2--Mstar}.
\item 
Medium-sized $Q$-balls with mass-to-charge ratio in the range
\beq 
3 \GeV \lesssim \frac{1}{\qX} \frac{M_Q}{Q} < m_\rm{NLSP} \sim \ms  \ .  \eeq
As per \eqs{eq:GMSB--M/Q} and \eqref{eq:GMSB--Q num}, this occurs for
\beq
70 \ < \ \qX \, b^\frac{d-2}{d-1} \ \lesssim \ 10^4 \pare{ \frac{\ms}{500 \GeV} }  \ ,
\label{eq:b medQ}
\eeq
and $Q$-ball charges in the interval
\beq
5 \cdot 10^{22} \pare{\frac{\mg}{1 \GeV}}^2 \pare{\frac{500 \GeV}{\ms}}^4  \frac{1}{\qX^4} \  
< \ Q \  \lesssim  
Q_\rm{stable}  \  .
\label{eq:Q medQ}
\eeq

The decay of $Q$-balls in this regime can produce only $R$-parity even particles and gravitinos. (Some NLSPs may also be produced at the late stages of $Q$-ball decay, when their charge has decreased and the mass-to-charge ratio has exceeded $m_\rm{NLSP}$.) 
This scenario has been studied in Ref.~\cite{Kasuya:2011ix}, where it was shown that the decay into gravitinos is suppressed, and that the gravitino abundance resulting from $Q$-ball decays is subdominant if the ellipticity of the condensate is $\e > 10^{-6} \pare{\ms \Mm/10^{12} \GeV^2 } \simeq 10^{-3} (\mg / 1 \GeV)$.
For the parameter $A$ of \eq{eq:A GMSB}, the ellipticity is given by \eq{eq:ellipt b>1}, and always satisfies this bound. 

Thus, for the range of $b$ given in \eq{eq:b medQ}, the only constraint from $Q$-ball decay we will consider is $T_Q \gtrsim 10 \MeV$, as in \eq{eq:TQ limit}. Using \eq{eq:GMSB--T b>1}, this limit becomes
\beq
b^{\frac{5(d-2)}{2(d-1)}} \pare{\frac{\mg}{1 \GeV}} \ < \ 10^5 \, f_Q^\frac{1}{2} \pare{\frac{\ms}{500 \GeV}}^\frac{5}{2}  \ .
\label{eq:Qdec con b>1 intermed}
\eeq
\item 
Small $Q$-balls with mass-to-charge ratio
\beq \frac{M_Q}{Q} > m_\rm{NLSP} \sim \ms \ . \eeq
As per \eqs{eq:GMSB--M/Q} and \eqref{eq:GMSB--Q num}, this occurs for 
\beq b^\frac{d-2}{d-1} \lesssim 70 \frac{1}{\qX} \ , \eeq
or
\beq Q \lesssim 5 \cdot 10^{22} \pare{\frac{\mg}{1 \GeV}}^2 \pare{\frac{500 \GeV}{\ms}}^4  \frac{1}{\qX^4} \ . \eeq
In this case, $Q$-balls can decay into $R$-parity odd particles (besides the gravitino), in addition to particles that are $R$-parity even. This is similar to the $b \lesssim 1$ case, and we shall adopt the same constraint, $T_Q \gtrsim \mg / 20$, as in \eq{eq:mg/20}. Given \eq{eq:GMSB--T b>1}, this becomes
\beq
b^{\frac{5(d-2)}{2(d-1)}} \pare{\frac{\mg}{1 \GeV}}^2  < 2 \cdot 10^4  \: f_Q^\frac{1}{2} \pare{\frac{\ms}{500 \GeV}}^\frac{5}{2}  \  .
\label{eq:Qdec con b>1 small}
\eeq
\end{list}

\bigskip

In summary, for the scenario of pangenesis explored in this paper, and for the GMSB case, we shall consider the parameter region bounded by the following relations:
\begin{align}
b^\frac{d-2}{d-1} \ 
&\lesssim \ 10^4 \pare{ \frac{\ms}{500 \GeV} } \frac{1}{\qX}  \ , \label{eq:summ Qdec}\\
b^{\frac{5(d-2)}{2(d-1)}} \pare{\frac{\mg}{1 \GeV}} \ 
&< \ 10^5 \: f_Q^\frac{1}{2} \pare{\frac{\ms}{500 \GeV}}^\frac{5}{2}  \ , \label{eq:summ Qdec>10MeV} \\
b^{\frac{5(d-2)}{2(d-1)}} \pare{\frac{\mg}{1 \GeV}}^2 \ 
&< \ 2 \cdot 10^4  \: f_Q^\frac{1}{2}  \pare{\frac{\ms}{500 \GeV}}^\frac{5}{2}  \ , \quad \rm{for} \ 1 <b \lesssim (70/\qX)^\frac{d-1}{d-2}  \ , \label{eq:summ b>1} \\
b \pare{\frac{\mg}{1 \GeV}}^2 \ 
&< \ 5\cdot 10^3  \: f_Q^\frac{1}{2} \pare{\frac{0.01}{|K|}}^\frac{1}{2}  \pare{\frac{\ms}{500 \GeV}}^\frac{5}{2}  \ , \quad \rm{for} \ b \lesssim 1 \ \rm{and \ if} \ K<0 \ .  \label{eq:summ b<1}
\end{align}
These boundaries arise respectively from: 
the opposite regime of \eq{eq:b bigQ}, ensuring $Q$-ball decay;
\eq{eq:Qdec con b>1 intermed}, ensuring $Q$-ball decay before BBN, at $T_Q \gtrsim 10 \MeV$;
and \eqs{eq:Qdec con b>1 small} and \eqref{eq:Qdec con b<1}, ensuring no overproduction of gravitinos from $Q$-ball decay.
We sketch these regions in Fig.~\ref{fig:m3/2--Mstar}, together with the constraints on the reheating temperature given in \eqref{eq:TR limit}.
We emphasise that at least some of the above bounds can be relaxed, according to the preceding discussion.

\section{Models of Pangenesis}
\label{sec:models}

In this section, we present a minimal dark sector which allows for successful pangenesis and discuss its phenomenology. 
In subsections \ref{sec:VS} and \ref{sec:DS}, we describe the particle content of the visible and dark sectors.
We identify suitable FDs for pangenesis in subsection \ref{sec:FDs}. 
In subsection \ref{sec:stability}, we discuss the stability of particles in association to the symmetries of the model. 
We describe the cascade of the anti-baryonic asymmetry to the lightest dark-sector particles in subsection \ref{sec:DarkAtoms}.

\subsection{The visible sector}
\label{sec:VS}

We model the visible sector by the MSSM, which contains the chiral superfields $Q$, $L$, $u^c$, $d^c$ and $e^c$ in the standard notation defined earlier. To ensure the cancellation of $(B-L)$-anomalies, we also include left-handed antineutrinos $n^c$. At the renormalisable level, the symmetries then allow for the superpotential 
\beq
W_\v  =  H_u Q u^c \, + \, H_d Q d^c \, + \, H_u L n^c \, + \, H_d L e^c \, + \, \mu H_u H_d\, ,
\label{eq:Wv}
\eeq
where, for simplicity, we have suppressed family indices and coupling constants, and omitted obvious SU(3)$_c$ and SU(2)$_L$ index contractions. 
Note that dangerous $B_\v$-violating terms are automatically absent due to $(B-L)$-invariance and that there is thus no need to impose an $R$-parity.

\subsection{A simple dark sector}
\label{sec:DS}

\begin{table}[t]
\begin{center}
 \begin{tabular}{r|r|r}
 & $B-L$ & $D$  \\
\hline
$\Delta $ & $q_\dm$ & 1 \\
\hline
$\Lambda $ & $0$ & -1\\
\end{tabular}
\caption{Charge assignments of dark sector fields. The partners $\bar{\Delta}$ and $\bar{\Lambda}$ carry the opposite charges. \label{table:partcont}} 
\end{center}
\end{table}

The dark sector must contain at least one chiral multiplet $\Delta$ which is charged under $B-L$. 
To ensure anomaly cancellation and allow for a supersymmetric mass term, we also include a partner $\bar{\Delta}$ with opposite charge assignments. 
In order for Eq.~\eqref{eq:asym rel} to establish a relation between VM and DM, we have to ensure that the symmetric part of the DM can annihilate efficiently to massless (or very light) particles.  This requires an annihilation cross-section somewhat larger than the weak scale \cite{Graesser:2011wi}. 
The prospect of efficient annihilation of DM into SM states are thus severely limited~\cite{Bai:2010hh,Goodman:2010ku}.
The interactions between the dark and visible sectors -- due to the $B-L$ vector multiplet and higher-dimensional operators -- are constrained to be suppressed by a larger scale, and give considerably smaller annihilation cross-sections of the DM to the visible sector\footnote{If the gauge-interaction which establishes the connection between the visible and the dark baryonic asymmetries couples to the visible sector as $B-xL$ with $x<1$, the annihilation cross-section of DM into SM fermions could be significantly larger than in the case of $B-L$. The collider constraints on a gauged $B-xL$ get significantly relaxed as $x\to 0$~\cite{Williams:2011qb}. Such an interaction would of course require the existence of exotic fermions to cancel the anomalies. \label{foot:ZB}}. 
This motivates introducing an unbroken gauge symmetry in the dark sector under which the dark relic baryons are charged, allowing the symmetric part of DM to annihilate to the dark force carriers.\footnote{Alternatively, the DM could annihilate to light dark fermions via Yukawa couplings.} The resulting annihilation cross section of the DM is $\sigma \sim \aD^2 m_\dm^{-2}$, where $m_\dm$ is the DM mass and $\aD$ the fine-structure constant of the dark gauge force. As we shall discuss in Sec.~\ref{sec:DS constraints} (see in particular Eq.~\eqref{eq:aD Bullet}), cosmological constraints on our model require that $\smash{\aD \gtrsim 0.1}$. Furthermore, the dark matter mass in pangenesis is predicted to be $m_\dm=\mathcal{O}$(GeV) (see Eq.~\eqref{eq:mDM gen} in Sec.~\ref{sec:DarkAtoms}). Together, this guarantees a sufficiently large annihilation cross section of the DM. 

We shall consider the simplest choice for this gauge symmetry, an abelian U(1)$_D$, under which $\Delta$ and $\bar{\Delta}$ are charged. 
In order to allow for U(1)$_D$-invariant monomials with nonvanishing $(B-L)$-charge (which combine with visible-sector monomials to form suitable FDs as we discuss in the next section), we also introduce two chiral multiplets $\Lambda$ and $\bar{\Lambda}$ with equal and opposite charges under U(1)$_D$. For definiteness, we shall choose the latter multiplets to carry vanishing $(B-L)$-charge, whereas we assign $(B-L)$-charges $\pm q_\dm$ to $\Delta$ and $\bar{\Delta}$, respectively. Our conclusions will not depend on this choice, though, as only the $(B-L)$-charges of the U(1)$_D$-invariant monomials $\Delta \Lambda$ and $\bar{\Delta}\bar{\Lambda}$ are relevant for the phenomenology of this model. The charge assignments are summarised in Table \ref{table:partcont} and allow for the superpotential
\beq
W_\d \,= \,  m_\delta \, \Delta \bar{\Delta} \, + \, m_\lambda \, \Lambda \bar{\Lambda} \,. 
\label{eq:dsp}
\eeq
The masses $m_\delta$ and $m_\lambda$ are taken to be around the GeV scale. We outline how this mass scale may be generated dynamically in Sec.~\ref{dsGMSB} and will present a more detailed model elsewhere \cite{wip}. 

The Weyl fermions in the supermultiplets $\Delta$ and $ \bar{\Delta}$ combine to form a Dirac fermion $\delta$ with mass $m_\delta$. Similarly, another Dirac fermion $\lambda$ with mass $m_\lambda$ stems from $\Lambda$ and $\bar{\Lambda}$. We will denote the scalar superpartners of these fermions by the same symbols as the supermultiplets, i.e.~$\Delta,\bar{\Delta},\Lambda,\bar{\Lambda}$, and the U(1)$_D$-gaugino by $\lambda_D$. We collectively refer to the latter particles as the dark-sector superpartners.

\subsection{Combined flat directions}
\label{sec:FDs}

\begin{table}[t]
\centering
 \begin{tabular}{r|r|r}
& $B-L$ & $F$-flat?  \\
\hline
$n^c$ &  1 & $\checkmark$\\
\hline
$L H_u$ & -1 & \\
\hline
$u^c d^c d^c$ & -1 & $\checkmark$\\
\hline
$L L e^c$ & -1 & $\checkmark$\\
\hline
$Q L d^c $ & -1 & $\checkmark$\\
\hline
$Q Q Q H_d$ & 1 & \\
\hline
$Q H_d u^c e^c$ & 1 &\\
\hline
$d^c d^c d^c L L $ & -3 & $\checkmark$\\
\hline
$u^c u^c u^c e^c e^c$ & 1 & $\checkmark$\\
\hline
$(Q u^c) (Q u^c) e^c$ & 1 & $\checkmark$\\
\hline
$(Q Q Q)  Q u^c$ & 1 & $\checkmark$\\
\hline
$d^c d^c d^c L H_d $ & -2 & \\
\end{tabular}
\caption{Monomials up to 5th order in visible-sector fields which carry $B-L$ but are otherwise singlets (taken from \cite{Gherghetta:1995dv}). \label{table:monomials}} 
\end{table}

We shall now discuss how the visible and the dark sector particle content of the previous sections allows for flat directions in the scalar potential along which pangenesis can be implemented. 

Directions in field space are `flat' if both the $D$-term and $F$-term contributions to the scalar potential vanish. Identifying FDs is simplified by the following correspondence \cite{Buccella:1982nx}: Every gauge-invariant holomorphic monomial in the chiral superfields gives rise to a direction in field space along which the $D$-terms cancel. Let us consider a simple example: The $D$-term which belongs to the U(1)$_D$ vector multiplet reads
\beq
D_{_D} = |\Delta|^2 - |\bar{\Delta}|^2 - |\Lambda|^2 + |\bar{\Lambda}|^2 \, .
\eeq
Along the direction $\Delta =\bar{\Delta}=0$ and $\Lambda =\bar{\Lambda}=\phi$ in field space, where the scalar $\phi$ parametrizes the direction, this $D$-term vanishes identically. Since $\Lambda$ and $\bar{\Lambda}$ carry charge only under U(1)$_D$, all other $D$-terms vanish similarly if we set the remaining scalar fields to zero. This direction, which we associate with the monomial $\Lambda \bar{\Lambda}$, is thus `$D$-flat'.

The $F$-terms lift some of the $D$-flat directions in the field space. Efficient asymmetry generation via the AD mechanism requires the $F$-term contribution to vanish at least at the renormalisable regime.\footnote{In fact, as recent progress in the AD dynamics has shown, for successful asymmetry generation, the directions should not be lifted by terms with dimension $d\leqslant4$ in the superpotential (see Sec.~\ref{sec:AD}).} This imposes additional constraints in determining the flat directions of the scalar potential. 
For our simple example, the direction $\Lambda \bar{\Lambda}$, this requirement is in fact not satisfied due to the mass term in the superpotential in Eq.~\eqref{eq:dsp}. For the purpose of the AD mechanism, however, terms of positive mass dimension in the scalar potential are not problematic since they are suppressed by powers of the mass parameter over the field VEV in the early universe compared to quartic terms. (Such terms arise anyway from SUSY-breaking.) We will therefore still call such directions `flat'. A given FD is typically lifted by higher-dimensional operators in the superpotential. The FD associated with $\Lambda \bar{\Lambda}$, for example, is lifted by the operator $W_\rm{lift} \supset (\Lambda \bar{\Lambda})^2/M_*$.

Let us now identify FDs which are suitable for pangenesis, for the model specified in the previous two sections.
At the renormalisable level, $B-L$ is separately conserved in the visible and the dark sector, producing the accidental symmetries $\BLv$ and $\Bd$, respectively.\footnote{In the visible sector, $B_\v$ and $L_\v$ are in turn separately conserved, as noted earlier and will be discussed further below.} Pangenesis is effective along FDs which carry charge under $X = \BLv + \Bd $ but which are singlets under the orthogonal combination $B-L$ (see~Sec.~\ref{sec:pan}). Since the chiral multiplets of our model are charged either under $\BLv$ \emph{or} under $\Bd$, such FDs necessarily involve fields from both sectors.\footnote{The possibility of `connector fields' which carry both charges was considered in \cite{Bell:2011tn}.} 
We denote by $\mathcal{O}^{(\v)}_q$ monomials of visible-sector fields which carry $(B-L)$-charge $q$ but which are singlets with respect to all other gauge symmetries. Such monomials, up to 5th order in the fields, are listed in Table~\ref{table:monomials}~\cite{Gherghetta:1995dv}. Similarly, we denote corresponding monomials in dark-sector fields by $\mathcal{O}_q^{(\d)}$. The set of such monomials can be straightforwardly constructed using Table~\ref{table:partcont}. For $q\ne 0$ and to lowest order, these are $\D\L$ and $\bar{\D}\bar{\L}$. FDs which carry charge under $X$ can be parameterised by products of these two types of monomials:
\beq
\mathcal{O}_q^{(\d)}  \cdot \mathcal{O}_{-q}^{(\v)}  \, .                                                                                                                                                                                                                                                                                                                                                                                                                                                                                                                                                                                                                                                                                                                                                                                                                                                               
\label{combinedFD}
\eeq
Gauge-invariance of this product ensures the cancellation of $D$-terms along the associated direction in field space.

The constraints from $F$-flatness have been systematically studied for the MSSM in Ref.~\cite{Gherghetta:1995dv}. Visible-sector monomials $\mathcal{O}^{(\v)}_q$ which fulfil these constraints at the renormalisable level are marked as `$F$-flat' in Table~\ref{table:monomials}. 
Note that Ref.~\cite{Gherghetta:1995dv} considered the MSSM  without right-handed neutrinos. 
Including these fields yields additional $F$-term constraints which lift FDs involving the SU(2)$_L$-invariant combination $L H_u$ already at the renormalisable level. Furthermore, the product of different visible-sector monomials which individually correspond to $F$-flat directions does not always correspond to an $F$-flat direction itself.
This should be taken into account when using Table~\ref{table:monomials}. Let us now consider monomials $\mathcal{O}^{(\d)}_q$. Since the dark-sector superpotential in Eq.~\eqref{eq:dsp} only contains mass terms, all such monomials fulfil the constraints from $F$-flatness at the renormalisable level.

In summary, FDs which are suitable for pangenesis are associated with monomials of the type shown in Eq.~\eqref{combinedFD} with the constraint that the visible-sector component satisfies the $F$-flatness conditions arising from the renormalisable superpotential of \eq{eq:Wv}. The set of these monomials obviously depends on the $(B-L)$-charge of the `basis monomials' $\D\L$ and $\bar{\D}\bar{\L}$. 
For example, for the choices $\smash{q_\dm = \frac{1}{2},2,3}$, such directions are, amongst others:
\begin{align}
(\Delta  \Lambda)^2  u^c d^c d^c    \label{FD1/2} \quad  &\text{for $q_\dm = \frac{1}{2}$} , \\
\Delta  \Lambda \, (u^c d^c d^c)^2  \label{FD2}   \quad  &\text{for $q_\dm = 2$} ,\\
\Delta  \Lambda \, d^c d^c d^c L L  \label{FD3}   \quad  &\text{for $q_\dm = 3$} .
\end{align}
Since there are three families of quarks and leptons, each of these monomials actually gives rise to several FDs. 
The above monomials are (by construction) the lowest order gauge-invariant operators involving a particular set of fields. They ultimately lift the corresponding FDs, and they are responsible for the generation of a net $X$ charge, according to the dynamics described in the previous section.

However, it is possible that other non-renormalisable operators of lower order also contribute to the potential along the field directions associated with the monomials of pangenesis. This is the case if there exist gauge-invariant operators of the type
\beq
W_\rm{lift} \supset \frac{1}{\Mnr^{n-3}} \: \x \: \cal{O}_{n-1}(\f_i) \ ,
\label{eq:evil}
\eeq
where $\x$ is a field that does not participate in the FD under consideration, and $\cal{O}_{n-1}(\f_i)$ is a dimension-$(n-1)$ operator involving only FD fields $\f_i$ (but not necessarily the entirety of them). To reproduce the dynamics described in Sec.~\ref{sec:AD}, it is necessary that $n>d$, where $d$ is the dimension of the $X$-violating monomial in the superpotential. This ensures that the term $|\f|^{2(n-1)}$, which the superpotential term of \eq{eq:evil} contributes to the FD potential, has a negligible effect on the dynamics of the AD field.

In the preceding discussion, we identified FDs for which no operators of the type of \eq{eq:evil} with $n\leqslant3$ exist. 
Let us now discuss operators of this type, potentially relevant for directions of pangenesis, with $n>3$.
There are no such operators involving only dark-sector fields, as long as the FD involves either the monomial $\D \L$ or $\bar{\D} \bar{\L}$, but not both.
Indeed, all dark-sector gauge-invariant monomials have the form
\beq
W_\rm{lift} \supset \frac{1}{M_*^{2k +2\ell -3}} (\Delta \bar{\Delta})^k (\Lambda \bar{\Lambda})^\ell \, , 
\eeq
where $k$ and $\ell$ are positive integers. The contribution of these terms to the potential vanishes for $k+\ell \geqslant 2$ since $\bar{\Delta}=\bar{\Lambda}=0$, along a FD involving only $\D \L$ (and vice versa if the FD contains $\bar{\D} \bar{\L}$). On the other hand, the richer field content of the visible sector may allow for operators of the type of \eq{eq:evil}, If such operators exist, they may reduce the multiplicity of the FDs under consideration, or even lift them completely at a lower order than the monomials characterising these FDs.
A systematic study of all non-renormalisable gauge-invariant operators is beyond the scope of this work.\footnote{For the MSSM, the operators which lift a given FD have been systematically determined in \cite{Gherghetta:1995dv}. However, since our model includes right-handed neutrinos and $(B-L)$-invariance (which changes the set of higher-dimensional operators allowed by the symmetries), their analysis is not directly applicable.} 
We emphasise however that certain non-renormalisable visible-sector operators allowed by gauge invariance must be highly suppressed in order to satisfy bounds on the proton lifetime (we discuss this further in Sec.~\ref{sec:spPMSB}). In the following, we will thus ignore non-renormalizable visible-sector operators altogether, as their existence depends on unknown physics, and their absence can be accordingly justified. In this case, the monomials of the type in \eq{combinedFD} are the leading terms in the superpotential which lift the associated FDs. These operators provide the required violation of $X$ and $CP$ for successful pangenesis.

\subsection{Stable particles}
\label{sec:stability}

\subsubsection{Gravity-mediated SUSY breaking}
\label{sec:spPMSB}

We will now determine the stable and metastable particles in the visible and the dark sector, beginning with the case of gravity-mediated SUSY breaking. All scalars and gauginos (and the gravitino) then obtain masses of order the soft scale $\ms$. For simplicity, we shall focus our discussion on the case of integer $(B-L)$-charge assignments $q_\dm$ for the dark-sector multiplets. It turns out that the set of stable particles then falls into two different classes, depending on whether $q_\dm$ is even or odd. We shall first present the analysis for the case of even $q_\dm$ and point out the differences for odd charge assignments at the end of this section.

Let us first ignore interactions between the visible and the dark sector. Imposing $(B-L)$-invariance in the visible sector results in two U(1) symmetries, $L_\v$ and $B_\v$, with the linear combination $(B + L)_\v$ being an accidental symmetry (albeit violated by sphalerons). These symmetries stabilise, respectively, the lightest neutrino and the proton. In addition, the visible sector has a gauged U(1)$_{\rm{EM}}$ which keeps the lightest electrically charged particle -- the electron -- from decaying. All other gauge symmetries are either broken or confined. Fermion number conservation stabilises an additional particle. In order to identify this particle, it is convenient to absorb a discrete subgroup of $(B-L)_\v$ into the fermion number and define the $R$-parity of a given particle as usual by
\beq
P_R^{(\v)}  \equiv  (-1)^{3(B-L)_\v + 2 s}\, ,
\eeq
where $s$ is the spin. Note that this symmetry is not imposed but follows from $(B-L)_\v$-invariance and fermion number conservation. As usual, all SM particles carry even $R$-parity and all superpartners are odd. The lightest superpartner is thus stable. We will refer to this particle as the visible-sector LSP.

The particle content and the superpotential of the dark sector allow for two U(1) symmetries, which we can identify with $B_\d$ and U(1)$_D$. The `dark baryon number' $B_\d$ guarantees the stability of $\delta$ since it is the lightest particle charged under this symmetry. In analogy to the visible sector, we will therefore call $\delta$ the `dark proton'. The gauge symmetry $U(1)_D$ in turn stabilises $\lambda$ to which we will correspondingly refer as the `dark electron'. In absence of gauge interactions, the superpotential also has four abelian continuous $R$-symmetries. These symmetries are broken to a single $\rm{U}(1)$ $R$-symmetry by the coupling to gauginos (and the gravitino) which in turn is broken to a $\mathbb{Z}_2$ $R$-parity by gaugino masses and $A$-terms. Choosing vanishing U(1)$_R$-charge for $\delta$ and $\lambda$, these particles are even under the $R$-parity whereas the dark-sector superpartners are odd. This symmetry thus stabilises the lightest superpartner, which we shall call the dark-sector LSP. The $R$-parity of a given particle can be written as
\beq
P_R^{(\d)}  \equiv  (-1)^{-3 B_\d + D +2 s} \, ,
\eeq
where $D$ is the charge under U(1)$_D$. We thus see that this symmetry is a subgroup of U(1)$_D$ and fermion number conservation (the factor $(-1)^{-3 B_\d}$ was included for later convenience but does not affect the $R$-parity assignments since all dark-sector particles have, by assumption, even $B-L$ and thus even $B_\d$).

Let us now reintroduce interactions between the visible and the dark sector. These break the two separately conserved fermion numbers (or equivalently $R$-parities) in each sector down to one universally conserved quantum number. Using the fact that $\smash{B-L = (B - L)_\v - B_\d}$, we can combine the two symmetries into a universal $R$-parity:
\beq
P_R^{(\rm{gen})} \equiv  (-1)^{3(B-L) + D + 2 s}\, .
\label{generalizedR}
\eeq
This symmetry stabilises the overall LSP, i.e.~the lightest particle among the visible-sector and the dark-sector LSP. The heavier particle of the two, on the other hand, decays via the $(B-L)$-gaugino or the gravitino, which couple to both sectors. For example, if the overall LSP resides in the dark sector, the visible-sector LSP can decay via an off-shell $(B-L)$-gaugino: 
\beq
\text{LSP}_\v \rightarrow \lambda^*_{B-L} + \text{SM particles} \rightarrow \text{LSP}_\d + \text{dark-sector fermions} + \text{SM particles} \, .
\eeq
The decay rate is suppressed by the large mass $M_{B-L}$ of the $(B-L)$-gaugino. In order not to endanger BBN, we have to ensure that the lifetime is less than $0.01 \text{ s}$ (see e.g.~\cite{Kawasaki:2004qu}). This gives the constraint
\beq
\MBL \lesssim  10^{15} \sqrt{f_s} \left( \frac{\ms}{500 \text{ GeV}} \right)^{3/2}  \text{ GeV}\, ,
\label{B-L-bound}
\eeq
were $f_s$ is a phase-space factor (e.g.~$f_s \approx 10^{-3}$ for a three-body decay) and we have assumed that the gauge couplings of $B-L$ and U(1)$_D$ are of order 1. Note that decays between the two sectors via the gravitino are strongly suppressed due to its small coupling strength and are thus less important than the channel via the $(B-L)$-gaugino.

In addition, the superpotential contains higher-dimensional operators which couple the visible and dark sector. For example for $q_\dm=2$, the symmetries of the theory allow for the following $X$-violating terms to appear in the superpotential:
\begin{multline}
W \supset  \frac{1}{M_*}  \bar{\Delta} \bar{\Lambda} \pare{n^c}^2  +\frac{1}{M_*^3}  \bar{\Delta} \bar{\Lambda} \pare{n^c}^2 \bigl[\Delta \bar{\Delta} + \Lambda \bar{\Lambda} + H_u H_d \bigr] \\ + \frac{1}{M_*^3}  \Delta \Lambda  \pare{L H_u}^2 +  \frac{1}{M_*^4}\bar{\Delta}  \bar{\Lambda} \,  Q Q Q H_d n^c + \dots 
\label{hdo}
\end{multline}
Here the ellipsis refers to additional terms of order-7 and higher orders in the superfields and we have assumed that all $X$-violating terms arise at the same scale $M_*$ as the operators which lift the FDs suitable for pangenesis.
The couplings in Eq.~\eqref{hdo} break the separately conserved symmetries $B_\v$, $L_\v$ and $B_\d$ down to the global $B-L$. 
Since the lightest state charged under $B-L$ is the neutrino, we expect that the ordinary proton, as well as the dark proton become unstable. 

We shall first determine the lifetime of the ordinary proton which results for $q_\dm=2$ and derive a bound on the scale $M_*$ from the requirement that the proton is sufficiently stable. Since $B_\v$- and $L_\v$-violating operators for $q_\dm>2$ involve higher-dimensional monomials in visible-sector fields (cf.~Table~\ref{table:monomials}) and are thus suppressed by larger powers of $M_*$, this bound ensures sufficient stability of the proton in these cases as well.

The combination of the first and the last term of \eq{hdo} gives the lowest order decay mode of the proton to SM particles via intermediate dark-sector particles, for example via the decay chain:
\beq
p \rightarrow \bar{\Delta}^* + \bar{\Lambda}^* + \text{MSSM particles} \rightarrow \text{neutrinos} + \text{mesons} \, .
\label{eq:pdecaychain}
\eeq
This is the most dangerous channel involving the dark sector. As we will see in Sec.~\ref{sec:DarkAtoms}, all dark-sector particles are heavier than the proton and can thus only appear as off-shell states in the decay. Counting propagators of heavy scalars and gauginos, the decay rate of the proton resulting from Eq.~\eqref{hdo} can be estimated as
\beq
\Gamma  \sim  f_p \frac{m_p^{33}}{M_*^{10} \, \ms^{22}} \, ,
\eeq
where $m_p$ is the proton mass and $f_p \ll 1$ takes into account further suppression due to phase-space/loop factors and coupling constants. 
Requiring that the lifetime of the proton is larger than the observational bound, $\tau_p \gtrsim 10^{34}$ years, leads to a lower bound on the scale $M_*$:
\beq
M_*  \gtrsim  4  \GeV \cdot  \sqrt[10]{f_p}  \pare{\frac{500 \GeV}{\ms}}^{11/5} \, .
\label{Mstarbound}
\eeq 
This is always satisfied for the values of $M_*$ required to produce enough asymmetry. The bound is so mild because the decay proceeds via a dimension-8 \emph{and} a dimension-5 operator. 

Note that the operators $Q Q Q L$ and $u^c u^c d^c e^c$, which allow the proton to decay involving only the visible sector, are allowed by the symmetries.  
If these operators would arise at the scale $M_*$ similar to those in Eq.~\eqref{hdo}, the proton decay rate would be too large, for the values of $M_*$ we considered in Sec.~\ref{sec:AD}. In fact, these operators must be suppressed by a scale considerably larger than the Planck scale to allow for a sufficiently stable proton. A way to naturally obtain this suppression was presented e.g.~in \cite{Harnik:2004yp}. Since this issue already arises in the MSSM alone and is independent of our dark sector, we will not consider these operators further.

Similarly, the higher-dimensional operators make the dark proton $\delta$ unstable. The decay products include a dark anti-electron to ensure conservation of $D$-charge, and SM particles.\footnote{This is assuming that $m_\delta > m_\lambda$. In the opposite case, the dark anti-electron $\bar{\lambda}$ would become unstable. For very degenerate masses $m_\delta$ and $m_\lambda$, finally, both particles would be stable.}  
In Sec.~\ref{sec:DarkAtoms}, we will identify the dark proton with a component of the DM state. Accordingly, its lifetime should exceed the age of the universe, $\t_{_U} \sim 10^{10} \yr$. Depending on the decay products however, there may be more stringent bounds on the lifetime coming from measurements of the cosmic radiation backgrounds to which the decays contribute. 
We shall now derive a bound on $M_*$ from the requirement that the dark proton is sufficiently long-lived if $q_\dm=2$. In analogy to the case of the proton, this bound again guarantees sufficient stability for $q_\dm>2$ as well.

For $q_\dm=2$, the first term in Eq.~\eqref{hdo} allows the decay
\beq
\delta \rightarrow \bar{\lambda} + 2 \, \bar{\nu}_R  \,   ,
\label{eq:delta decay}
\eeq
at 1-loop via sneutrino and $(B-L)$-gaugino propagators. The decay rate is 
\beq
\Gamma  \sim  f_d \: \frac{\gBL^4}{16 \p^2} \frac{m_\delta^{13}}{M_*^2 \, \ms^8 \, \MBL^2} \, ,
\eeq
where $f_d \sim 10^{-3}$ contains the phase-space suppression of the 3-body final state.
Because the neutrino states produced in this decay are sterile, no bounds from radiation backgrounds apply. 
Requiring $\G^{-1} > 10^{10} \yr$ translates to a lower bound on $M_*$:
\beq
M_*  \gtrsim 4 \cdot 10^7 \GeV 
\pare{\frac{f_d}{10^{-3}}}^{1/2}  
\left(\frac{m_\delta}{3 \GeV}\right)^{13/2}
\left(\frac{500 \GeV}{\ms}\right)^4 
\pare{\frac{1 \TeV}{\MBL/\gBL^2}} \, .
\label{eq:dadecay}
\eeq
This is significantly stronger than the constraint of \eq{Mstarbound}, but is still always satisfied for the values of $M_*$ required for successful pangenesis (see Figs.~\ref{fig:PMSB TR vs Mnr} and \ref{fig:m3/2--Mstar}).

The dark-proton decay mode of \eq{eq:delta decay} is the channel with the largest decay rate since all other $B_\d$-violating operators are suppressed by larger powers of $M_*$. These higher-order operators induce DM decay into detectable SM particles (e.g. photons, charged leptons, active neutrinos), in which case constraints from cosmic radiation backgrounds apply. Even though these constraints on the lifetime of the dark proton are stronger than that imposed to derive \eq{eq:dadecay}, it turns out that the resulting bounds on $\Mnr$ are weaker than \eq{eq:dadecay}.

\bigskip

In the discussion so far, we have assumed that $q_\dm$ is even. Let us now consider the case of odd $(B-L)$-charge assignments in the dark sector. 
The dark proton $\delta$ then becomes an odd state with respect to the universal $R$-parity of Eq.~\eqref{generalizedR}. As the lightest $R$-parity odd particle, it is absolutely stable. The dark electron $\lambda$ remains stabilised by its charge under U(1)$_D$. However, the lightest superpartner (among both visible-sector and dark-sector particles) -- which for even $q_\dm$ was the lightest $R$-parity odd state and thus stable -- becomes unstable due to the higher-dimensional operators that break the global $\Bv, \ \Lv$ and $\Bd$ down to $B-L$. 
The assignment $q_\dm=1$ is not suitable for pangenesis, as it generically allows for the superpotential term $\bar{\Delta} \bar{\Lambda} \: n^c$ and breaks the $X$ symmetry at the renormalisable level. We shall thus consider the case $q_\dm=3$.

For $q_\dm=3$, the terms allowed in the superpotential to the two lowest orders are
\beq
W \supset  \frac{1}{M_*^2} \, \bar{\Delta} \bar{\Lambda} \, \pare{n^c}^3 + \frac{1}{M_*^4} \, \Delta \Lambda \, d^c d^c d^c L L \  .
\label{hdo2}
\eeq
These operators induce the decay of the lightest superpartner. For example, the first term allows the decay of the $D$-gaugino
\beq
\l_D 
\ \to \
\pare{ \delta + \bar{\D}^* }
\ \rm{or} \
\pare{ \l + \bar{\L}^* }
\ \to \
\delta + \l + 3 \, \nu_R    \ ,
\label{eq:lambdaD decay}
\eeq
where the off-shell $\bar{\D}$ or $\bar{\L}$ decays at 1-loop via sneutrino and $(B-L)$-gaugino propagators. The decay rate is estimated to be
\beq
\Gamma  \sim  f_r \, \frac{\gBL^4 g_{_D}^2}{16 \p^2} \, \frac{\ms^7}{M_*^4 \, \MBL^2} \, ,
\eeq
resulting in a lightest superpartner with lifetime
\beq
\t \sim 5 \cdot 10^{16} \yr
\pare{ \frac{\Mnr}{10^{13} \GeV} }^4
\pare{ \frac{0.1}{g_{_D}} }^2
\pare{ \frac{500 \GeV}{\ms} }^7
\pare{ \frac{\MBL/\gBL^2}{1 \TeV} }^2
\pare{ \frac{10^{-5}}{f_r} }   \,  .
\eeq
That is, for all the values of $\Mnr$ of interest for successful pangenesis (see Figs.~\ref{fig:PMSB TR vs Mnr} and \ref{fig:m3/2--Mstar}), the lifetime is considerably larger than the age of the universe, $\t_{_U} \sim 10^{10} \yr$.
In Sec.~\ref{sec:DarkAtoms}, we shall see under what conditions the lightest superpartner is a subdominant component of DM. Note that for the decay mode of \eq{eq:lambdaD decay}, no constraints from cosmic radiation backgrounds apply -- even if the lightest-superpartner abundance is comparable to that of DM -- since the only SM particles produced are sterile neutrino states. Higher-order operators produce decay modes with detectable SM final states, but the corresponding decay rates are suppressed by additional powers of $\ms/\Mnr$ and no constraints relevant for pangenesis arise.

The decay of the ordinary proton into visible-sector particles can proceed via the combination of dimension-8 and dimension-6 operators arising from the terms of \eq{hdo2}.  The resulting decay rate is then suppressed with respect to the $q_\dm=2$ case at least by a factor of $(m_p/\Mnr)^2$, and the lower bound on $\Mnr$ is even more relaxed than \eq{Mstarbound}. For odd charges $q_\dm > 3$, the bounds are further relaxed.

\subsubsection{Gauge-mediated SUSY breaking}
\label{dsGMSB}

Let us now consider the case of gauge-mediated SUSY breaking. The mass spectrum depends on the set of gauge groups under which the messengers are charged. 
In the simplest case, the messengers couple to all gauge groups in the visible and the dark sector. As usual, the gauginos then obtain masses of order the soft scale $\ms$ at 1-loop and the scalars at 2-loop. The same particles as in the gravity-mediated case are stable with the exception of the LSP which is now the gravitino. The bound on the $(B-L)$-breaking scale $\MBL$ in Eq.~\eqref{B-L-bound}, that we imposed to ensure that NLSPs decay sufficiently fast to LSPs, accordingly no longer applies. It is replaced by the bound on the reheating temperature for large gravitino masses that we have discussed in Sec.~\ref{sec:GMSB--grav}.

Another possibility is that the messengers are charged under the SM gauge groups and U(1)$_{B-L}$ but not under U(1)$_D$. The visible-sector and $B-L$ gauginos again obtain masses at 1-loop and all scalars at 2-loop (the dark-sector scalars via $(B-L)$-gauginos). Naively, one expects that the mass of the U(1)$_D$-gaugino $\lambda_D$ is generated via 3-loop diagrams (with intermediate $(B-L)$-gauginos and dark-sector chiral multiplets).
It was shown in \cite{ArkaniHamed:1998kj}, however, that the mass in this situation actually only arises at 5-loop order and can be estimated as
\beq
m_{\lambda_D}  \sim  \frac{g_{_{D}}^6 \gBL^4}{(8 \pi^2)^5} \, \ms \sim  g_{_{D}}^6  \gBL^4 \left(\frac{\ms}{500 \text{ GeV}}\right) \cdot 100 \text{ eV} \, .
\eeq
Here $g_{_{D}}$ and $\gBL$ are, respectively, the gauge couplings of U(1)$_D$ and U(1)$_{B-L}$. It is thus possible that $\lambda_D$ replaces the gravitino as the LSP. This may ease the constraints on the reheating temperature from gravitino overclosure and NLSP decays that we have discussed in Sec.~\ref{sec:GMSB--grav}. Since the coupling of $\lambda_D$ to dark-sector particles necessarily involves heavy scalars, it decouples at temperatures below the soft scale and is thus a hot thermal relic. To avoid overclosure of the universe, $m_{\lambda_D}$ must therefore be smaller than $100 \text{ eV}$ which is readily achieved.

Finally, let us briefly mention the case that the messengers couple to the SM gauge groups but to neither U(1)$_{B-L}$ nor U(1)$_D$. Scalar mass-squareds $m_{ds}^2$ in the dark sector are then generated via intermediate SM fields and $(B-L)$-gauginos at 4-loop, leading to 
\beq
m_{ds}^2 \sim \frac{\ms^2}{(16 \pi^2)^2} \sim \left(\frac{\ms}{500 \text{ GeV}}\right)^2 (3 \text{ GeV})^2 \,.
\label{eq:mDS=GeV}
\eeq
This case is thus promising for the dynamical generation of the GeV mass scale for the dark matter in pangenesis. We will present such a model, where this mass is induced via the dark-sector soft scale $m_{ds}$, elsewhere \cite{wip}.

\subsection{Dark atoms as dark matter}
\label{sec:DarkAtoms}

Pangenesis generates an asymmetry in the dark baryon number $B_\d$, which is carried by the components of the $\D, \bar{\D}$ multiplets. 
Since the dark gauge symmetry U(1)$_D$ remains unbroken, a compensating asymmetry resides in the $\L, \bar{\L}$ multiplets. 

The scalar components of the chiral multiplets decouple, decay and transfer the asymmetry into the fermionic degrees of freedom, according to
\begin{align}
\D &\to \ \delta + \l_D  \qquad  \rm{or} \qquad   \D \to \ \delta + \l_{B-L}^*  \  \to \ \delta + \rm{LSP}_\v + \rm{SM \ fermions}  \label{eq:Delta decay} \\
\L &\to \ \l     + \l_D  \qquad  \rm{or} \qquad   \L \to \ \l     + \l_{B-L}^*  \  \to \ \l     + \rm{LSP}_\v + \rm{SM \ fermions}  \label{eq:Lambda decay}
\end{align}
and similarly for $\bar{\D}, \: \bar{\L}$. 
The decay modes on the left assume that the scalars are heavier than the U(1)$_D$ gaugino, $m_\D, m_\L > m_{\l_D}$. 
If this is not the case, the dark-sector scalars can still decay via an off-shell $\l_{B-L}$, as shown on the right.
Of course, the decay of the dark-sector scalars presupposes that none of them is the overall LSP.
This is essential for the success of our scenario in the generic case when the transmission of SUSY breaking to the dark sector is unsuppressed. We will return to this point later in this section.

The dark baryonic asymmetry eventually resides in the dark protons $\delta$, which combine with the asymmetric abundance of the dark electrons $\l$ to form $D$-neutral dark atoms. The latter constitute the DM of the universe in our scenario. 
After  $\delta, \: \l$ become non-relativistic, their symmetric thermal populations annihilate away into $\ZD$ bosons.
We shall now calculate the relic DM mass density and compare it with that of VM, using the structural relation of pangenesis, Eq.~\eqref{eq:asym rel}. 

The conserved  $\Bd$ and $D$ charge-to-entropy ratios are\footnote{We denote $\h_f = [ n(f) - n(\bar{f}) ] / s$, for a field $f$. Since we are concerned with particle populations whose symmetric part is annihilated away, this is also the particle-number-to-entropy ratio for that field, $\h_f = n(f)/s$ or $\h(f) = -n(\bar{f})/s$.}
\begin{align}
 \h[\Bd] &= q_\dm \: \h_\delta = \frac{\h[X]}{2} \ , \label{eq:eta Bd} \\
 \h[D]   &= \h_\delta - \h_\l = 0 \ ,  \label{eq:eta D}
\end{align}
where $q_\dm$ is the $X$ or, equivalently, $\Bd$ charge of the $\delta$ fermions (and of the dark atoms).
Equations \eqref{eq:eta Bd} and \eqref{eq:eta D} give
\beq
\h_\delta = \h_\l = \frac{1}{q_\dm} \frac{\h[X]}{2} \ .
\label{eq:eta delta lambda}
\eeq
Note that \eq{eq:eta delta lambda}, and thus the relations that follow, would still hold for any $\Bd$-charge assignment for the fields $\Delta, \ \Lambda$ which maintained $q_\delta + q_\lambda = q_\dm$. In fact, shifting the baryonic charges of $\Delta, \ \Lambda$ by equal and opposite amounts is equivalent to kinetic mixing of U(1)$_{B-L}$ with U(1)$_D$.

In the visible sector, the number density of protons over entropy density $\eta_p$ is 
\beq
\h_p = \eta[\Bv]  =  a_s \:  \eta[\BLv]  =  a_s \:  \frac{\eta[X]}{2} \, ,
\eeq
where the factor $a_s$ takes into account the possible processing of the $\BLv$ asymmetry by electroweak sphalerons. If the asymmetry is released in the thermal bath before the electroweak phase transition, then ${a_s \simeq 0.35}$ \cite{Harvey:1990qw}. However, if the reheating temperature is lower than the electroweak scale, and/or the condensate fragments into $Q$-balls that decay after the electroweak phase transition, then $a_s = 1$.

Using the above, the ratio of the dark and visible matter abundances is
\beq
\frac{\Omega_\dm}{\Omega_\text{\tiny{VM}}} 
= \frac{m_\delta \: |\eta_\delta| + m_\l \: |\eta_\l| }{m_p \: |\eta_p|}  
=  \frac{(m_\delta + m_\lambda)/ q_\dm } {a_s \: m_p} 
\simeq 5 \, .
\eeq
Thus, the observed DM mass density is obtained if the mass of the dark atoms is
\beq
m_\dm  \simeq  5 \, q_\dm \: a_s \: m_p \simeq   q_\dm \cdot (1.6 - 5) \text{ GeV}\, , 
\label{eq:mDM gen}
\eeq
where $m_\dm \simeq m_\delta + m_\lambda$ (the binding energy of $\delta$ and $\lambda$ is negligible). 
The DM mass range of \eq{eq:mDM gen} and its scaling with the baryonic charge of DM is a general prediction of baryon-symmetric models, quite independently of the detailed structure of the dark sector.\footnote{If sphaleron-type effects exist in the dark sector, and process the $\Bd$ asymmetry, the DM mass prediction should be modified by an appropriate factor.}
Equation~\eqref{eq:mDM gen} also shows that as long as $q_\dm \geq 1$, the proton is kinematically disallowed to decay into the dark sector, as assumed earlier in this section.

We now return to the point made earlier, that the decay of the dark-sector scalars is necessary for the success of our scenario: Assume that one of the scalars, e.g. $\D$, is the overall LSP. Then, inverse decays of the type $\D + \bar{\delta} \to \l_D$ decouple at $T \simeq m_{\l_D}/2 > m_\D /2$, while $\D$ is still relativistic. At this point, $\D$ carries a significant portion of the baryonic asymmetry, shared via chemical equilibrium with the fermionic degrees of freedom: $\h_\D = \h_\delta$. After the decoupling of the gaugino inverse decays, the individual U(1)$_R$ symmetries of each chiral multiplet are restored (see Sec.~\ref{sec:stability}). This means that $\h_\D$ will be conserved (independently of $\h_\delta$), and will contribute to the relic DM abundance (the symmetric part of the $\D$ particles annihilates into massless $\ZD$ bosons). If $m_\D \sim \ms \sim 10^2 \GeV$, \eq{eq:mDM gen} cannot be satisfied (unless $q_\dm \sim 10^2$), and the $\D$ relic abundance would overclose the universe. We note, however, that if the SUSY-breaking transmission to the dark sector is suppressed and produces soft masses of order $m_{ds} \sim \text{GeV}$ (see \eq{eq:mDS=GeV}), a dark-scalar LSP can provide a viable scenario of pangenesis.

\bigskip

In order to allow the dark atoms to constitute the dominant part of DM, we have to ensure that the LSP is underabundant. Its freeze-out abundance can be sufficiently small if the LSP resides in the visible sector and is mostly wino or mostly higgsino~\cite{Mizuta:1992ja,Mizuta:1992qp,Edsjo:1997bg,Drees:1996pk}.
Alternatively, the LSP may be the U(1)$_D$ gaugino, which annihilates into the dark-sector fermions $\delta$ or $\l$ via a scalar exchange of $\D$ or $\L$, respectively. 
The cross-section for this process is $\s_\rm{ann} \approx \aD^2/m_{ds}^2$. Cosmological constraints set $\aD \gtrsim 0.1$ (see \eq{eq:aD Bullet}), which provides for a larger annihilation cross-section than that of a visible WIMP LSP. Additional enhancement of the LSP annihilation arises if the dark-sector soft scale is suppressed compared to that in the visible sector, $m_{ds} \ll \ms$ (in which case the LSP naturally resides in the dark sector).

\section{Cosmological constraints on dark sectors}
\label{sec:DS constraints}

We now discuss constraints on our dark sector arising from BBN, large scale structure formation and the Bullet cluster.

\subsection{Dark radiation and BBN}
\label{sec:BBN}

A key feature of asymmetric DM scenarios, including those of pangenesis, is the annihilation of the symmetric part of the dark matter plasma into dark radiation.  This gives rise to a constraint from BBN, because the additional radiation from the dark sector will increase the expansion rate of the universe during this epoch and potentially endanger the successful prediction of the primordial light element abundances.  

It has become customary to quantify an additional radiative component of the thermal bath during BBN by an ``effective number of extra neutrino species'', $\delta N_\rm{eff}$.  Currently, data on the primordial abundances allows extra radiation equivalent to about one additional neutrino species: $\delta N_\rm{eff} \le 1$ at 95\% C.L.~\cite{Mangano:2011ar}.  This additional radiation is consistent with WMAP determinations of the relativistic energy density at recombination, which stands at $\delta N_\rm{eff} = 1.34 ^{+0.86}_{-0.88}$ at 68\% C.L.~\cite{Komatsu:2010fb}. It thus suffices to impose the constraint $\delta N_\rm{eff} \le 1$.

The radiative energy density due to relativistic dark sector species is determined both by the number of such degrees of freedom, and by the temperature of the dark plasma.  The latter is not expected to be same as that of the visible sector for temperatures below the decoupling temperature $\Tdec$, the lowest temperature at which the two sectors are in thermal equilibrium with each other.  The histories of how entropy is released from massive species into radiation will, in general, be different in each sector.  For our purposes, we wish the dark sector to acquire a temperature $\Td$ that is sufficiently lower than that of the visible sector $\Tv$ in order to suppress the relativistic energy density contributed by dark radiation.

The evolution of the ratio $\Td/\Tv$ is easily tracked through entropy conservation, which implies that
\beq
\frac{\gv(\Tv) \Tv^3}{\gd(\Td) \Td^3} = \frac{\gv(\Tdec)}{\gd(\Tdec)} \ ,
\eeq
where $\gv(T)$ and $\gd(T)$ are the temperature-dependent degree-of-freedom counts in the visible and dark sectors, respectively, and $\Tdec$ is the common temperature at the time the two sectors decouple.  
Using this, the $\delta N_\rm{eff} \le 1$ limit becomes
\beq
\gd(\Tdec) \lesssim 36 \pare{\frac{\gdN}{2}}^{1/4} \pare{\frac{\gv(\Tdec)}{213.5}} \ .
\eeq

The visible and the dark sectors are kept in equilibrium via the gauged $B-L$ interaction. They decouple at temperatures around the $B-L$ breaking scale, $\Tdec \sim \MBL\gtrsim M_\rm{EW}$. Thus, assuming MSSM content for the visible sector, $\gv(\Tdec) \approx 213.5$.
For the dark sector of Sec.~\ref{sec:models}, $\gd(\Tdec) = 20$, and $\gdN = 2$ at the onset of BBN ($\Tv \simeq 1$ MeV), so the $\delta N_\rm{eff} \le 1$ requirement is fulfilled.

\subsection{Dark-matter recombination and self-interaction}
\label{sec:self-int}

Our dark sector furnishes `atomic' DM, meaning that the DM today is actually a hydrogen-like bound state of fundamental dark fermions due to the unbroken U(1)$_D$  gauge interaction which acts like an electromagnetic force in that sector.  This kind of DM has now been well studied (see e.g.~\cite{Kaplan:2009de,Petraki:2011mv}), and we simply quote previous results and explain how they apply to the present scenario.

The first constraint arises from the need to have successful large scale structure (LSS) formation.  To ensure an appropriately fast growth of structure in the universe, overdensities driven by gravitational attraction must develop in the dark sector sufficiently early.  From general considerations, it is well known that structure will only begin to form when the universe becomes matter-dominated.  But density inhomogeneities also cannot develop before the recombination of the dark atoms, because the growth of gravitational inhomogeneities will be disrupted by the relatively strong U(1)$_D$ force.  To ensure good LSS, it suffices, therefore, to require that the dark atoms recombine into $D$-neutral bound states no later than the point of matter-radiation equality.  The criterion for when recombination can be considered to be effectively complete is that the ionisation fraction is at most $0.1$. Requiring that the residual ionisation fraction after recombination is no more than this value leads to the constraint~\cite{Petraki:2011mv}
\beq
\aD \ge 3.5 \times 10^{-3} \left( \frac{10^{-9}}{\eta[\Bd]} \right)^{\frac{1}{4}} \left( \frac{\mD}{1\ \rm{GeV}} \right)^{\frac{1}{4}}\ ,
\eeq
while the necessity to have recombination complete before matter-radiation equality enforces~\cite{Petraki:2011mv}
\beq
\mD \gtrsim 0.5\, \rm{keV}\, \left( \frac{0.3}{\aD} \right)^2\ ,
\eeq
where $\mD$ is the reduced mass of the dark atom system, and $\aD$ is the U(1)$_D$ fine-structure constant. Both of these constraints are mild.

Finally, observations on the Bullet cluster place an upper bound on the scattering cross-section $\sigma$ for dark atom collisions with themselves, estimated to be~\cite{Markevitch:2003at,Randall:2007ph}
\beq
\frac{\sigma}{m_\dm} \lesssim 1\, \rm{cm}^2\, / \, \rm{g}.
\label{eq:BCconstraint}
\eeq
The cross-section is related to the dark atom properties through~\cite{Kaplan:2009de}
\beq
\sigma \simeq 4\pi k^2 a_o^2\ ,
\eeq
where $3 \lesssim k \lesssim 10$, and $a_o = ( \aD\, \mD )^{-1}$ is the Bohr radius.  Using this in \eq{eq:BCconstraint} leads to the bound
\beq
\aD \gtrsim 0.1\ \left( \frac{k}{3} \right) \left( \frac{1\ \rm{GeV}}{\mD} \right) \left( \frac{3.2\ \rm{GeV}}{m_\dm} \right)^{\frac{1}{2}}\ ,
\label{eq:aD Bullet}
\eeq
which is the most stringent of the constraints derived in this section.

\section{Signatures}
\label{sec:signatures}

\subsection{Collider signatures}
\label{sec:collider signatures}

The generalised and gauged $B-L$ interaction is the most generic signature of pangenesis. 
Collider experiments currently constrain $\ZBL$ to be heavier than about $500 \GeV$ for $\smash{\aBL =\gBL^2/4\p \sim 10^{-2}}$~\cite{Williams:2011qb}. A massive $\ZBL$ may thus be within the reach of the LHC. If produced, it is possible to measure its invisible decay~\cite{Emam:2007dy,Basso:2009hf,Basso:2008iv,Basso:2010pe,Basso:2010yz,Coutinho:2011xb,
Leike:1998wr,Rizzo:2006nw,Langacker:2008yv,Petriello:2008zr,Coriano:2008wf,Petriello:2008pu,Gershtein:2008bf,Langacker:2009im}. If the latter cannot be accounted for by neutrinos, such a measurement will be significant evidence for a baryonic dark sector.

\subsection{Dark-matter direct detection}

Antibaryonic dark matter can be detected via $\ZBL$ exchange. 
The spin-independent scattering cross section per nucleon can be as high as
\beq
\s_{_{B-L}}^\rm{SI}    \approx  
8 \cdot 10^{-41} \cm^2  
\ \ \gBL^4 \:
\pare{ \frac{1 \TeV}{\MBL} }^4 \ ,
\label{eq:sigmaDD_B-L}
\eeq
where we used $m_\dm = 5 \GeV$. Current direct-detection experiments are not sensitive to such low masses, and there are therefore no constraints from XENON100~\cite{Aprile:2011hi} or CDMS II~\cite{Ahmed:2010wy}. However, models of pangenesis can easily accommodate larger masses, if the baryonic charge of the lightest $\Bd$-charged particle is larger (cf. \eq{eq:mDM gen}). 
We note that the direct-detection cross section may be higher if the gauge interaction which establishes the connection between the visible and the dark asymmetries couples to the visible sector as $B-xL$ with $x<1$ (see footnote \ref{foot:ZB}).

It is also possible that the dark U(1)$_D$ force mixes kinetically with the photon. If it is broken, the DM state is the dark proton alone, and thus carries $D$ charge. The kinetic mixing between the dark photon and the ordinary photon provides another direct-detection channel with cross section
\beq
\s_{_D}^\rm{SI}    \approx   10^{-36} \cm^2 
\pare{ \frac{\e}{10^{-4}} }^2      
\pare{ \frac{g_{_D}}{0.1} }^2     
\pare{ \frac{100 \MeV}{M_{_D}} }^4 ,
\label{eq:sigmaDD_dark}
\eeq
where we used  $m_\dm = 5 \GeV$.
Such an interaction could account for the regions favoured by DAMA and CoGeNT~\cite{Savage:2008er,Aalseth:2010vx}.

We note in passing that a broken U(1)$_D$ can support a successful dark-sector scenario for pangenesis if the $\ZD$ has sub-GeV mass, such that the DM particles can annihilate into dark photons. In order for the symmetric part of DM to be annihilated  efficiently via U(1)$_D$, the gauge coupling has to be $g_{_D} \gtrsim 0.05$. 
The massive dark gauge bosons $Z_{_D}'$ can decay into lepton pairs of the visible sector via the kinetic mixing to the photon. For $\ZD$ mass about  $100 \MeV$, the kinetic mixing can be $\e \sim 10^{-4}$~\cite{Bjorken:2009mm,Dent:2012mx}. This allows for a rapid decay of the $\ZD$ bosons immediately after they decouple. No extra relativistic energy density is then present at BBN, and no long-range force among DM particles exists, which in turn means no astrophysical constraints. The kinetic mixing between the dark force and the photon can be probed in current fixed target experiments~\cite{Bjorken:2009mm,Essig:2010xa}.

\section{Conclusions}
\label{sec:conc} 

The similar present-day mass densities for ordinary and dark matter suggest that they have a common origin in the early universe, and that the DM density is due to a particle-antiparticle asymmetry that matches the situation in the visible sector.  A common cosmological origin for both types of matter is a key feature of baryon-symmetric models, which see the universe as being symmetric under a generalised baryon number that counts both ordinary baryons and dark matter particles.  In these scenarios, an antibaryon asymmetry that exactly cancels the visible baryon asymmetry\footnote{Modulo sphaleron reprocessing.} is hidden from view in the dark matter of the universe.  This enforces a tight relationship between the number densities of VM and DM and helps to explain their observed similarity.  The DM mass scale is then predicted to be in the few-GeV regime.

In this paper we reported on a detailed analysis of one of the main ways of producing a baryon-symmetric universe: the mechanism of pangenesis, which uses Affleck-Dine dynamics.  We focussed on all relevant aspects of both gravity-mediated and gauge-mediated supersymmetry breaking: asymmetry generation, gravitino production, and $Q$-ball formation and decay.  We showed that viable regions of parameter space exist for both cases, and when considering gauge-mediation we left the messenger scale arbitrary.  The main results are summarised in Figs.~\ref{fig:PMSB TR vs Mnr} and \ref{fig:m3/2--Mstar}.

We also presented a class of viable dark sectors.  Typical features include GeV-scale masses and a dark-sector analogue of electromagnetism that is required for annihilating away the symmetric part of the dark plasma.  The dark matter today is then typically predicted to be `atomic', meaning that stable DM particles form hydrogen-like bound states due to their Coulomb attraction.  Constraints from big bang nucleosynthesis, large scale structure formation and the Bullet cluster were derived, leaving a viable region of parameter space.

There are two important non-gravitational interactions between the VM and DM in pangenesis.  The most characteristic feature is a $Z'$ gauge boson that couples to both sectors.  The simplest models see the $Z'$ coupling to ordinary matter via $B-L$.  This particle can be produced in accelerators such as the LHC, but it will have a substantial invisible width into dark-sector species, distinguishing it from standard $Z'$ possibilities.  The discovery of such a $Z'$ would obviously be a spectacular advance in understanding DM and its putative relationship to VM.  Kinetic mixing between the ordinary photon and the dark photon provides another potentially important interaction between the sectors, and can be probed in direct DM detection experiments.  Overall, the following would constitute evidence for pangenesis: supersymmetry, together with GeV-scale DM mass and a $Z'$ having an invisible width that cannot be accounted for by neutrinos.

\section*{Acknowledgements}
We would like to thank Nicole Bell for collaboration during the early stages of this work.
We also thank Alex Kusenko for very useful comments, and James Barnard, Anupam Mazumdar and Ian Shoemaker for helpful discussions.
This work was supported, in part, by the Australian Research Council.

\bibliographystyle{JHEP}
\bibliography{Bibliography.bib}

\end{document}